\documentclass[aps,prb,twocolumn,showpacs]{revtex4-1}

\usepackage{graphicx}
\usepackage{amsmath}
\usepackage{color}

\newcommand{\bfG}{{\bf G}}
\newcommand{\bfq}{{\bf q}}

\begin{document}

\title{Assessment of long-range-corrected exchange-correlation kernels for solids:\\
accurate exciton binding energies via an empirically scaled Bootstrap kernel}

\author{Young-Moo Byun}
\affiliation{Department of Physics and Astronomy, University of Missouri, Columbia, MO 65211, USA}

\author{Carsten A. Ullrich}
\email[]{ullrichc@missouri.edu}
\affiliation{Department of Physics and Astronomy, University of Missouri, Columbia, MO 65211, USA}

\begin{abstract}
In time-dependent density-functional theory, a family of exchange-correlation kernels,
known as long-range-corrected (LRC) kernels, have shown promise in the calculation of excitonic effects
in solids. We perform a systematic assessment of existing static LRC kernels
(empirical LRC, Bootstrap, and jellium-with-a-gap model)
for a range of semiconductors and insulators, focusing on optical spectra and exciton binding energies.
We find that no LRC kernel is capable of simultaneously producing good optical spectra
and quantitatively accurate exciton binding energies for both semiconductors and insulators.
We propose a simple and universal, empirically scaled Bootstrap kernel which yields accurate
exciton binding energies for all materials under consideration, with low computational cost.
\end{abstract}


\maketitle

\section{Introduction} \label{sec:I}

The optical properties of insulators and semiconductors in
the energy range close to the gap are strongly influenced by excitons. The accurate and efficient calculation of excitonic properties is
an important task of computational materials science, since it is a key requirement in the
design of novel photovoltaic materials of desired properties. For example, low exciton binding energies in perovskite solar cells promote the electron-hole separation and thereby enhance power conversion efficiencies.\cite{Miyata2015}

Many-body perturbation theory is a standard theoretical method for excitonic effects in solids: accurate exciton binding energies $E_{\mathrm{b}}$ and optical absorption spectra of semiconductors and insulators are obtained by solving the Bethe-Salpeter equation (BSE).\cite{Rohlfing98,Onida2002,Reining2016} However, the BSE is computationally too expensive to be applied to large systems.
Time-dependent density-functional theory (TDDFT)\cite{Ullrich2012,Ullrich2015} provides alternatives to the BSE which are computationally much cheaper.

The main challenge for TDDFT lies in finding approximations to the exchange-correlation (xc) kernel $f_{\mathrm{xc}}$ which
yield accurate excitonic properties. The random-phase approximation (RPA) ($f_{\mathrm{xc}} = 0$), the local-density approximation (LDA), and generalized gradient approximations (GGAs) fail to capture excitonic effects in solids due to their inadequate long-range behavior. The so-called ``nanoquanta kernel'',\cite{Reining02,SOR03,ADM03,MDR03} constructed by reverse-engineering the BSE,
yields very good optical spectra of solids and thus provides an important proof of principle; however, it is computationally as
expensive as the BSE.

Hybrid xc functionals (mixtures of semilocal xc functionals with a fraction
of nonlocal Fock exchange) are very widely used in TDDFT. The B3LYP hybrid functional \cite{SDCF94} gives
reasonably good optical spectra for systems whose gap is not too large.\cite{Bernasconi2011,Tomic2014} For organic molecular crystals,
the so-called optimally tuned range-separated hybrids produce excellent results.\cite{Refaely2015}
A scaled exact exchange approach was recently shown to yield good excitonic binding energies for a wide variety of
materials.\cite{Yang2015} However, the nonlocal exchange contribution adds to the computational cost of the hybrid methods;
it is therefore desirable to work with purely local xc functionals.

A simple nonlocal model kernel, which is known as the long-range-corrected (LRC) kernel,\cite{Reining02,Botti04,Botti2007}
\begin{equation}
f_{\mathrm{xc}}^{\mathrm{LRC}} = -\frac{\alpha}{\mathbf{q}^2}, \label{eq:LRC}
\end{equation}
where $\bfq$ is the momentum transfer in the first Brillouin zone (BZ), can account for bound excitons in solids, but it requires a material-dependent parameter $\alpha$, a positive scalar.
A number of xc kernels proposed in the literature, such as the empirical LRC, Bootstrap, RPA-Bootstrap, and jellium-with-gap-model (JGM) kernels,
\cite{Botti04,Botti2007,Sharma11,Rigamonti15,Trevisanutto13} report that the long-range part in them gives the most important contribution to their results, and
we hence refer to them as the family of LRC-type kernels.
These kernels have been applied to simple bulk semiconductors and
insulators, with some degree of success. However, there also were reports of conflicting results, giving rise to some
recent controversies in the literature.\cite{Sharma16,Rigamonti16}

Testing the performance of the various LRC-type kernels is a complex task which depends on many choices.
For instance, the xc kernel, which is formally a matrix in reciprocal space, can be implemented
as head-only, diagonal, or a full matrix. Local-field effects can be fully or partially included, or completely ignored.
The calculated optical spectra depend on the input band structure (LDA with or without scissors correction, GGA,
LDA+U, hybrids, or GW) and on the method (such as all-electron versus pseudopotential-based). And, last but not least, the selection of the
materials is important. Given the large number of choices that have to be made, an unbiased assessment
and a comparison between different LRC methods is challenging, and conflicting results can arise.

In this paper, we will perform a systematic assessment of the various existing \emph{static} LRC-type kernels (i.e. we do not assess \emph{dynamical} LRC-type kernels such as those proposed in Refs.~\onlinecite{Botti05,Berger15}), for a variety of materials ranging
from small-gap semiconductors to large-gap insulators, comparing calculated optical spectra and exciton binding energies to
experimental data. The main finding is that the existing LRC-type kernels, while often producing good-looking optical spectra for
semiconductors, all fail to yield consistently good exciton binding energies. We propose an empirical scaling approach,
to be used in conjunction with the RPA-Bootstrap method, which gives accurate $E_{\rm b}$ for all materials under study,
but the resulting optical spectra may have unsatisfactory distributions of oscillator strength.

This paper is organized as follows. In Section \ref{sec:II}, we give an overview of the formal framework of linear-response TDDFT, comparing
two approaches to describe optical properties of solids: the Dyson-equation approach and the Casida equation.
We then review the existing static LRC-type xc kernels, and the different choices for their implementation. We also discuss some
computational details. Section \ref{sec:III} then presents our results. We demonstrate the sensitivity of the optical spectra to
the choice of the $\alpha$-parameter, and then propose a scaled RPA-Bootstrap kernel which gives accurate exciton binding energies.
Section \ref{sec:IV} contains our conclusions.

\section{Background and methodology} \label{sec:II}

\subsection{Linear-response TDDFT for solids: Dyson equation vs Casida equation approach} \label{subsec:IIA}

There are several ways to calculate optical absorption spectra of periodic systems using linear-response TDDFT.\cite{Ullrich2015}
The most common approach is based on the interacting density-density response function $\chi_{\bfG\bfG'}(\mathbf{q}, \omega)$,
where $\bfG$ and $\bfG'$ are reciprocal lattice vectors, and $\omega$ is the frequency. The response function is obtained from the following Dyson-type equation:
\begin{align}
&\chi_{\bfG\bfG'}(\bfq, \omega)=
\chi^{(0)}_{\bfG\bfG'}(\mathbf{q}, \omega)+
\sum_{\bfG_1\bfG_2}\chi^{(0)}_{\bfG\bfG_1}(\mathbf{q}, \omega) \nonumber \\
&\times \left[V_{\bfG_1}(\mathbf{q})\delta_{\bfG_1\bfG_2}
+
f_{\mathrm{xc},\bfG_1\bfG_2}(\mathbf{q})\right]\chi_{\bfG_2\bfG'}(\mathbf{q}, \omega),
\label{dyson}
\end{align}
where $\chi^{(0)}$ is the noninteracting response function and
$V_\bfG(\bfq)=4\pi/|\mathbf{q}+\mathbf{G}|^{2}$ is the Coulomb interaction. It is convenient to write
$V = V_0 + \bar V$, where $V_{0}$ is the long-range ($\mathbf{G}=0$) part of the Coulomb interaction, and $\bar{V}$ is the Coulomb interaction without the long-range part. $f_{{\rm xc},\bfG\bfG'}(\mathbf{q})$ is the xc kernel in the adiabatic approximation,
i.e.,  independent of $\omega$. $\chi^{(0)}$ is explicitly given by~\cite{Gajdos06}
\begin{align}
&\chi_{\mathbf{GG'}}^{(0)}(\mathbf{q},\omega) = \frac{2}{\mathcal{V}} \sum_{nm\mathbf{k}} (f_{m\mathbf{k+q}}-f_{n\mathbf{k}}) \nonumber \\
&\times \frac{\langle m\mathbf{k+q}|e^{i(\mathbf{k+G})\cdot\mathbf{r}}|n\mathbf{k} \rangle \langle n\mathbf{k}|e^{-i(\mathbf{k+G'})\cdot\mathbf{r'}}|m\mathbf{k+q} \rangle}{E_{m\mathbf{k+q}}-E_{n\mathbf{k}}-(\omega+i\eta)}, \label{eq:chi0}
\end{align}
where $\mathbf{k}$ lies within the first BZ, $n$ and $m$ are band indices, $E_{n\mathbf{k}}$ and $E_{m\mathbf{k+q}}$ are the associated Kohn-Sham single-particle energies, $f = 1(0)$ for occupied (unoccupied) states, the factor of 2 accounts for the spin (we here only consider
non-spin-polarized systems), $\mathcal{V}$ is the crystal volume, and $\eta$ is an infinitesimal. In the optical limit ($\mathbf{q} \to 0$), the head ($\mathbf{G}=\mathbf{G'}=0$) of $\chi^{(0)}$ at $\omega = 0$ becomes\cite{Baroni86}
\begin{align}
\chi^{(0)}_{00}(\mathbf{q}\to0,0)&=-\frac{4\mathbf{q}^{2}}{\mathcal{V}}\sum_{vc\mathbf{k}}\frac{|\langle c\mathbf{k}|\hat{p}+i[V_{\mathrm{NL}},\hat{r}]|v\mathbf{k}\rangle|^{2}}{(E_{c\mathbf{k}}-E_{v\mathbf{k}})^{3}} \label{eq:chi0.00}
\end{align}
where $v$ and $c$ are valence and conduction band indices, respectively, $\hat{p}$ is the momentum operator, $\hat{r}$ is the position operator, and $V_{\mathrm{NL}}$ is the non-local part of the pseudopotential.
The $\bfq^2$-dependence will be important for the construction of the Bootstrap kernels, see below. It is also important that $\chi^{(0)}_{00}(\mathbf{q}\to0,0)$ is always negative.

The optical spectrum is obtained from the imaginary part of the macroscopic dielectric function $\epsilon_{\mathrm{M}}$:
\begin{align}
\epsilon_{\mathrm{M}}(\omega)&=\lim_{\mathbf{q} \to 0} \frac{1}{\epsilon_{00}^{-1}(\mathbf{q},\omega)} \\
&=\lim_{\mathbf{q} \to 0} \frac{1}{1+V_{0}(\mathbf{q})\chi_{00}(\mathbf{q},\omega)}, \label{eq:dielectric.function}
\end{align}
where $\epsilon^{-1}$ is the inverse dielectric function.\cite{Onida2002} We shall refer to this method as the
Dyson approach; it has a moderate computational cost, and is therefore the method of choice for calculating optical spectra.
However, the drawback of the Dyson-equation approach is that fine details of the spectra, in particular the binding energies
of weakly bound excitons, cannot be obtained, because the spectral broadening washes out any subtle features of the order of a few meV
(see also Sec. \ref{subsec:IIC}).

As an alternative which is strictly equivalent to the Dyson equation, optical spectra and exciton binding energies can be
obtained from the Casida equation:\cite{Casida1995}
\begin{equation}
 \begin{pmatrix}
  \mathbf{A} & \mathbf{B} \\
  \mathbf{B}^{*} & \mathbf{A}^{*}
 \end{pmatrix}
 \begin{pmatrix}
  X_{n} \\
  Y_{n}
 \end{pmatrix}
 =
 \omega_{n}
 \begin{pmatrix}
  \mathbf{-1} & \mathbf{0} \\
   \mathbf{0} & \mathbf{1}
 \end{pmatrix}
 \begin{pmatrix}
  X_{n} \\
  Y_{n}
 \end{pmatrix}, \label{eq:Casida}
\end{equation}
where $\mathbf{A}$ and $\mathbf{B}$ are excitation and de-excitation matrices, respectively, $X_{n}$ and $Y_{n}$ are $n$th eigenvectors, and $\omega_{n}$ is the $n$th excitation energy.
%\textcolor{red}{Note that Dyson and Casida equations are in principle strictly equivalent.}
The matrix elements of $\mathbf{A}$ and $\mathbf{B}$ are given by
\begin{align}
A_{vc\mathbf{k},v'c'\mathbf{k'}}&=(E_{c\mathbf{k}}-E_{v\mathbf{k}})\delta_{vv'}\delta_{cc'}\delta_{\mathbf{k}\mathbf{k'}}
+F^{\mathrm{Hxc}}_{vc\mathbf{k},v'c'\mathbf{k'}}, \\
B_{vc\mathbf{k},v'c'\mathbf{k'}} &=F^{\mathrm{Hxc}}_{vc\mathbf{k},v'c'\mathbf{k'}},
\end{align}
where $F^{\mathrm{Hxc}}=F^{\mathrm{H}}+F^{\mathrm{xc}}$ is the Hartree-exchange-correlation (Hxc) matrix.\cite{Ullrich2012}
In the optical limit, $F^{\mathrm{H}}$ and $F^{\mathrm{xc}}$ are given by
\begin{align}
F^{\mathrm{H}}_{vc\mathbf{k},v'c'\mathbf{k'}}
&=
\frac{2}{\cal V} \sum_{\mathbf{G}\neq0} \frac{4\pi}{|\mathbf{G}|^{2}} \langle c\mathbf{k}|e^{i\mathbf{G}\cdot\mathbf{r}}|v\mathbf{k}\rangle \langle v'\mathbf{k'}|e^{-i\mathbf{G}\cdot\mathbf{r}}|c'\mathbf{k'}\rangle, \label{eq:FH} \\
F^{\mathrm{xc}}_{vc\mathbf{k},v'c'\mathbf{k'}}
&=
\frac{2}{\cal V}\lim_{\bfq\to 0}
\sum_{\mathbf{GG'}} f_{\mathrm{xc},\mathbf{GG'}}(\bfq) \langle c\mathbf{k}|e^{i(\bfq + \mathbf{G})\cdot\mathbf{r}}|v\mathbf{k}\rangle
\nonumber\\
&
\times \langle v'\mathbf{k'}|e^{-i(\bfq + \mathbf{G}')\cdot\mathbf{r}}|c'\mathbf{k'}\rangle. \label{eq:Fxc}
\end{align}
For the elements of $F^{\mathrm{xc}}$ in Eq.~(\ref{eq:Fxc}) to remain finite (i.e., neither vanishing nor diverging) in the $\mathbf{q} \to 0$ limit,
the head ($\bfG=\bfG'=0$) of $f_{\mathrm{xc}}$ should be proportional to $\mathbf{q}^{-2}$,
the wings ($\bfG=0,\bfG'\ne0$ or vice versa)  should be proportional to $\mathbf{q}^{-1}$,
and the body ($\bfG,\bfG'\ne 0$)  should be independent of $\mathbf{q}$. In other words, the most general form is
\begin{align}   \label{fxcmatrix}
\lim_{\bfq\to 0} f_{{\rm xc}, \bfG\bfG'}(\bfq)=
 \begin{pmatrix}
  \frac{\kappa_{00}}{\mathbf{q}^2} & \frac{\kappa_{01}}{\mathbf{q}} & \frac{\kappa_{02}}{\mathbf{q}} & \cdots \\
  \frac{\kappa_{10}}{\mathbf{q}} & \kappa_{11} & \kappa_{12} & \cdots \\
  \frac{\kappa_{20}}{\mathbf{q}} & \kappa_{21} & \kappa_{22} & \cdots \\
  \vdots & \vdots & \vdots & \ddots
 \end{pmatrix}
 ,
\end{align}
where the $\kappa_{\mathbf{GG'}}$ are constants (in general, they are functionals of the density).
We will discuss various approximations of the xc kernel  in the following subsection.

The excitation energy spectrum $\omega_n$ of the Casida equation (\ref{eq:Casida}) for periodic solids with a gap has discrete levels,
which correspond to bound excitons, and a continuous part, which corresponds to the unbound particle-hole excitations. For the
adiabatic xc kernels considered here, only one excitonic level is found, which can be identified as the lowest bound exciton
(to obtain an excitonic Rydberg series with a scalar xc kernel requires the kernel to be frequency-dependent).\cite{Yang2012,Yang2013}
We calculate the exciton binding energy as that energy which separates this discrete level from the onset of the continuum.
Since no artificial spectral broadening is involved, exciton binding energies can be calculated in principle with arbitrary precision.
However, the Casida-equation approach is computationally expensive because it requires building and diagonalizing a large matrix.

Note that by using a very small broadening width and a very fine frequency grid, one may be able to obtain $E_{\mathrm{b}}$ of semiconductors from the Dyson-equation optical spectrum, but the broadening width and the frequency grid spacing always cause an error that may be greater than $E_{\mathrm{b}}$ of interest. Note also that Ref.~\onlinecite{Rigamonti15} proposed a method to ``read'' $E_{\mathrm{b}}$ from the real part of $\epsilon_{\mathrm{M}}^{\mathrm{RPA}}(\omega)$, but this approach works only for head-only kernels and only for wide-gap insulators (i.e. one cannot obtain small $E_{\mathrm{b}}$ on the order of a few meV), and it has a moderate precision ($\sim$0.1 eV). By contrast, the Casida equation works for all forms of the xc kernel and for all materials, and it has a high precision ($\sim$0.01 meV).

A widely used approach to simplify the Casida equation is the so-called
Tamm-Dancoff approximation (TDA), which decouples excitations and de-excitations by setting $\mathbf{B}$ to zero in Eq.~(\ref{eq:Casida}).
However, we have found \cite{Byun2017} that the TDA underestimates LRC $E_{\mathrm{b}}$ of insulators significantly (i.e. by more than 100\%) (e.g. TDA and full Casida equations using the RPA-Bootstrap kernel without the scissors shift yield $E_{\mathrm{b}}=666$ and 2400~meV, respectively, for solid Ne), so
we will only use the full Casida equation in this work.

The local-field effect (LFE) is determined by the number of $\mathbf{G}$ vectors included, and has different forms in the Dyson and Casida equations. In the Dyson approach, the LFE means including not only the head, but also the wings and body of the matrix in $\mathbf{G,G'}$, which leads to $\epsilon_{00} \neq 1/\epsilon^{-1}_{00}$.
The Dyson equation is used to calculate optical spectra and Bootstrap-type kernel parameters (more about this later). In the Dyson equation for optical spectra, the LFE is not a matter of choice and should be included. However, in the Dyson equation for Bootstrap-type kernel parameters, the LFE is a matter of choice because of the freedom of defining Bootstrap-type kernels.
In this work, we will include the LFE when calculating Bootstrap-type kernel parameters, following the convention adopted in the literature.\cite{Sharma11}

In the Casida equation, the LFE means including not only the head, but also other terms in the summation of $F^{\mathrm{Hxc}}$ matrix elements in Eqs.~(\ref{eq:Fxc}) and (\ref{eq:FHxc}). Mathematically, the LFE in the Dyson equation is exactly transformed into the summation in the Casida equation. Therefore, if the LFE is included in the Dyson equation, it should be included in the Casida equation, too.

\subsection{LRC-type xc kernels} \label{subsec:IIB}
In reciprocal space, the xc kernels $f_{{\rm xc},\bfG \bfG'}(\bfq)$ are matrices in $\bfG$ and $\bfG'$, see Eq. (\ref{fxcmatrix}). In the following,
we list the xc kernels we have tested, paying particular attention to distinguish between head-only, diagonal or full matrix
forms. In all expressions, the optical limit $(\bfq\to 0)$ is understood.

\subsubsection{Empirical LRC kernel} \label{subsubsec:IIB1}

The empirical LRC kernel was originally designed as a kernel for optical spectra of semiconductors.~\cite{Reining02}
The diagonal and the head-only versions of the empirical LRC kernel are defined, respectively, as
\begin{equation}
 f_{\mathrm{xc}}^{\rm LRC(d)}
 = -\frac{\alpha}{4\pi}V =
 \begin{pmatrix}
  -\frac{\alpha}{\mathbf{q}^2} & 0 & 0 & \cdots \\
  0 & -\frac{\alpha}{\mathbf{G}_{1}^2} & 0 & \cdots \\
  0 & 0 & -\frac{\alpha}{\mathbf{G}_{2}^2} & \cdots \\
  \vdots & \vdots & \vdots & \ddots
\end{pmatrix}
\end{equation}
and
\begin{equation} \label{LRCh}
 f_{\mathrm{xc}}^{\rm LRC(h)} = -\frac{\alpha}{4\pi}V_{0} =
\begin{pmatrix}
  -\frac{\alpha}{\mathbf{q}^2} & 0 & 0 & \cdots \\
  0 & 0 & 0 & \cdots \\
  0 & 0 & 0 & \cdots \\
  \vdots & \vdots & \vdots & \ddots
 \end{pmatrix}.
\end{equation}
Here, $\alpha$ is an empirical parameter, given by\cite{Botti04}
\begin{equation}
\alpha_{\mathrm{LRC}} = \frac{C_{1}}{\epsilon_{\infty}} - C_{2},
\end{equation}
where $C_{1}=4.615$, $C_{2}=0.213$, and $\epsilon_{\infty}$ is the high-frequency dielectric constant. Note that here we use $\epsilon_{\mathrm{RPA}}^{-1}$ instead of the experimental $1 / \epsilon_{\infty}$, where $\epsilon_{\mathrm{RPA}}^{-1}$ is greater than $1 / \epsilon_{\infty}$ by $\sim$10\%. Also note that the empirical LRC kernel used calculated lattice parameters, while we take experimental ones. Due to these differences, empirical parameters $C_{1}$ and $C_{2}$ should be re-fitted to our choices, but it turns out that such differences make little effect on LRC results for semiconductors (see below).

In general, when a head-only or diagonal LRC kernel is used, $F^{\mathrm{Hxc}}$ simplifies to
\begin{align}
F^{\mathrm{Hxc}}_{vc\mathbf{k},v'c'\mathbf{k'}}
=\frac{2}{\cal V} \Big( \sum_{\mathbf{G}\neq0} \frac{4\pi - \bar{\alpha}}{|\mathbf{G}|^{2}} \langle c\mathbf{k}|e^{i\mathbf{G}\cdot\mathbf{r}}|v\mathbf{k}\rangle \langle v'\mathbf{k'}|e^{-i\mathbf{G}\cdot\mathbf{r}}|c'\mathbf{k'}\rangle \nonumber \\
-\alpha_{0} \frac{\langle c\mathbf{k}|\hat{p}+i[V_{\mathrm{NL}},\hat{r}]|v\mathbf{k}\rangle}{E_{c\mathbf{k}}-E_{v\mathbf{k}}} \frac{\langle c'\mathbf{k'}|\hat{p}+i[V_{\mathrm{NL}},\hat{r}]|v'\mathbf{k'}\rangle^{*}}{E_{c'\mathbf{k'}}-E_{v'\mathbf{k'}}} \Big), \label{eq:FHxc}
\end{align}
where $\alpha=\alpha_{0}$ and $\bar{\alpha}=0$ for the head-only kernel $f_{\mathrm{xc}}^{\rm LRC(h)}$, and $\alpha=\alpha_{0}=\bar{\alpha}$ for the diagonal kernel $f_{\mathrm{xc}}^{\rm LRC(d)}$. Note that head-only or diagonal LRC kernels reduce the exciton Hamiltonian building time drastically because this removes the double loop
over $\mathbf{G, G'}$ in Eq.~(\ref{eq:Fxc}).

It turns out that the body of $f_{\mathrm{xc}}^{\rm LRC(d)}$ has a negligible effect on optical spectra of semiconductors such as Si:\cite{Botti04} this is because $\bar{\alpha} \approx 0.2 \ll 4\pi$ in Eq.~(\ref{eq:FHxc}). However, $f_{\mathrm{xc}}^{\rm LRC(h)}$ and
$f_{\mathrm{xc}}^{\rm LRC(d)}$ can produce very different results for insulators, and one needs to state clearly which version,
(h) or (d), of the xc kernel is used.

\subsubsection{Bootstrap kernels} \label{subsubsec:IIB2}

The original Bootstrap kernel is a parameter-free xc kernel for optical spectra of semiconductors and insulators.~\cite{Sharma11} The original Bootstrap kernel is defined as
\begin{equation}
f^{\mathrm{B}}_{\mathrm{xc},\mathbf{GG'}}(\mathbf{q},\omega) = \frac{V^{1/2}_{\mathbf{G}}(\mathbf{q}) \epsilon^{-1}_{\mathbf{GG'}}(\mathbf{q},0) V^{1/2}_{\mathbf{G'}}(\mathbf{q})}{1-\epsilon_{\mathrm{RPA},00}(\mathbf{q}, 0)}, \label{eq:fxc.fB}
\end{equation}
where $\epsilon^{-1}$ is the self-consistent (i.e. bootstrapped) inverse dielectric function.
In matrix form, the bootstrap kernel is given by
\begin{align} \label{fxcBoot}
 f_{\mathrm{xc}}^{\mathrm{B}} =
 \begin{pmatrix}
  \frac{\beta_{00}}{\mathbf{q}^2} & \frac{\beta_{01}}{|\mathbf{q}||\mathbf{G}_{1}|} & \frac{\beta_{02}}{|\mathbf{q}||\mathbf{G}_{2}|} & \cdots \\
  \frac{\beta_{10}}{|\mathbf{G}_{1}||\mathbf{q}|} & \frac{\beta_{11}}{\mathbf{G}_{1}^2} & \frac{\beta_{12}}{|\mathbf{G}_{1}||\mathbf{G}_{2}|} & \cdots \\
  \frac{\beta_{21}}{|\mathbf{G}_{2}||\mathbf{q}|} & \frac{\beta_{21}}{|\mathbf{G}_{2}||\mathbf{G}_{1}|} & \frac{\beta_{22}}{\mathbf{G}_{2}^2} & \cdots \\
  \vdots & \vdots & \vdots & \ddots
 \end{pmatrix}
 ,
\end{align}
where
\begin{align}
\beta_{\mathbf{GG'}} = \frac{4 \pi \epsilon^{-1}_{\mathbf{GG'}}(\mathbf{q}, 0)}{1 - \epsilon_{\mathrm{RPA},00}(\mathbf{q}, 0)}.
\end{align}
Neglecting the wings and body of $f_{\mathrm{xc}}^{\mathrm{B}}$, which can be viewed as neglecting the LFE, yields a head-only Bootstrap kernel:
\begin{align}\label{fxcBooth}
f_{\mathrm{xc}}^{\rm B(h)} =
 \begin{pmatrix}
  \frac{\beta_{00}}{\mathbf{q}^2} & 0 & 0 & \cdots \\
  0 & 0 & 0 & \cdots \\
  0 & 0 & 0 & \cdots \\
  \vdots & \vdots & \vdots & \ddots
\end{pmatrix} .
\end{align}
Comparing $f_{\mathrm{xc}}^{\rm B(h)}$ with $f_{\mathrm{xc}}^{\rm LRC(h)}$, we define the LRC $\alpha$-parameter for
the Bootstrap kernel as
\begin{equation}
\alpha_{\rm B} = \frac{4\pi \epsilon^{-1}_{00}(0,0)}{\epsilon_{\mathrm{RPA},00}(0,0) - 1}. \label{eq:fxc.hB}
\end{equation}
Whereas $f_{\mathrm{xc}}^{\rm LRC(d)}$ and $f_{\mathrm{xc}}^{\rm LRC(h)}$ give quite different results for insulators, we have found that $f_{\mathrm{xc}}^{\rm B}$ and $f_{\mathrm{xc}}^{\rm B(h)}$ make a relatively small difference for \emph{both}
semiconductors and insulators
(see Table~I in Supplemental Material\cite{supplemental}), which is consistent with the findings of Refs.~\onlinecite{Sharma11,Rigamonti15}.
In Ref.~\onlinecite{Trevisanutto13} the same trend was found in the JGM kernel (see below).
Therefore, in view of the reduced computational effort,
we use the head-only form for all kernels in the following unless stated otherwise. The only exception is when we verify the results of Dyson-equation optical spectra obtained from $f_{\mathrm{xc}}^{\mathrm{B}}$ using Casida-equation calculations
(see Table~I in Supplemental Material\cite{supplemental}).
We emphasize again that we only consider the $\bfq \to 0$ limit here; at finite $\bfq$, the matrix character of the
Bootstrap kernel appears to play a more significant role.\cite{Sharma2012}

We also consider two simpler variations of the Bootstrap kernel. The first one, referred to as the 0-Bootstrap
kernel,\cite{Sharma_commun} is the head-only Bootstrap kernel (\ref{fxcBooth}) without the built-in self-consistency (i.e.,
``0'' means no iteration, similar to the G$_{0}$W$_{0}$ version of the GW approach) for optical spectra of semiconductors and insulators.
The LRC $\alpha$-parameter for the 0-Bootstrap kernel is thus given by
\begin{equation}
\alpha_{\rm 0B} = \frac{4\pi \epsilon^{-1}_{\mathrm{RPA},00}(0,0)}{\epsilon_{\mathrm{RPA},00}(0,0) - 1}. \label{eq:fxc.0B}
\end{equation}
Note that $\alpha_{\mathrm{0B}} > \alpha_{\rm B}$ by about 10\% because $\epsilon_{\mathrm{RPA},00}^{-1}(0,0)$ is
greater than $\epsilon^{-1}_{00}(0,0)$ by about 10\%.

The second simplified Bootstrap kernel is the RPA-Bootstrap kernel,\cite{Rigamonti15} which is a head-only kernel with
\begin{equation}
\alpha_{\rm RPAB} = \frac{4\pi \epsilon^{-1}_{\mathrm{RPA},00}(0,0)}{1 / \epsilon^{-1}_{\mathrm{RPA},00}(0,0) - 1} \label{eq:fxc.RPAB}
\end{equation}
for exciton binding energies of insulators. Note that $\alpha_{\mathrm{RPAB}} > \alpha_{\mathrm{0B}}$ by about 10\% because $1 / \epsilon^{-1}_{\mathrm{RPA},00}(0,0) - 1 < \epsilon_{\mathrm{RPA},00}(0,0) - 1$ by about 10\%. Note also that without the LFE (i.e. when $\epsilon_{00}=1/\epsilon^{-1}_{00}$), the 0-Bootstrap and RPA-Bootstrap kernels become identical.

\subsubsection{Jellium with a gap model} \label{subsubsec:IIB3}

The JGM kernel is a parameter-free kernel for optical spectra of semiconductors and insulators.\cite{Trevisanutto13} The JGM kernel is defined as
\begin{eqnarray}
f^{\mathrm{JGM}}_{\mathrm{xc}}(\mathbf{q}; n, E_{\mathrm{g}})
&=& \frac{4\pi}{\mathbf{q}^{2}} \left(\frac{B(n) + E_{\mathrm{g}}}{1 + E_{\mathrm{g}}}\right)
\left[e^{k'_{n,E_{\mathrm{g}}}\mathbf{q}^2}-1\right]\nonumber\\
&-&
\frac{4\pi}{k^{2}_{\mathrm{F}}} \frac{\mathbf{q}^2}{(\mathbf{q}^2+1)} \frac{C(n)}{1 + E_{\mathrm{g}}}
\end{eqnarray}
with
\begin{equation}
k'_{n,E_{\mathrm{g}}} = k_{n} + \frac{E_{\rm g}^2}{4\pi n \mathbf{q}^{2}} \left( \frac{1 + E_{\mathrm{g}}}{B(n) + E_{\mathrm{g}}}\right) \:.
\end{equation}
Here, $E_{\mathrm{g}}$ is the band gap, $n$ is the electron density, and $k_{\mathrm{F}}$ is the Fermi wave vector; $k_{n}$, $B(n)$, and $C(n)$
are defined in Ref.~\onlinecite{Constantin07}. $f^{\mathrm{JGM}}_{\mathrm{xc},\mathbf{GG'}}(\mathbf{q}; E_{\mathrm{g}})$ is defined
as a full matrix, obtained from the Fourier transform in reciprocal space and the symmetrization in $\mathbf{G, G'}$; however,
we here use it in the head-only form. Whereas other LRC-type kernels depend
on dielectric constants, the JGM kernel depends on band gaps.

\subsection{Computational aspects} \label{subsec:IIC}

We used the Abinit code,\cite{Gonze09} which is based on norm-conserving pseudopotentials, for calculating the Kohn-Sham band structures including scissors
corrections, as well as GW band gaps within the LDA. Experimental lattice parameters were used for all materials.
We used the dp code\cite{Olevano97} for calculating optical spectra from the imaginary part of the dielectric function.
We calculated exciton binding energies from the Casida equation with our own homemade code.

Optical spectra were obtained with a Lorentzian broadening of 0.15 eV for GaAs, and 0.2 eV for all other materials. Note that a Lorentzian broadening smaller than 0.15 eV, which is an optimal value that makes calculated and experimental $E_{2}$ peaks have similar heights, generates an artificial $E_{1}$ peak in the excitonic region of the RPA and LRC spectra of GaAs,
and makes RPA and LRC $E_{2}$ peaks higher than the experimental one, and thus should not be used (see the top panel of Fig.~\ref{fig2} for $E_{1}$ and $E_{2}$ peaks). The Lorentzian broadening has a physical meaning (i.e. it simulates the lifetime broadening and can be calculated from GW),~\cite{Cazzaniga11} so it should not be used as an adjustable parameter to improve the appearance of calculated optical spectra.

We used experimental band gaps, $E_{\mathrm{g}}^{\mathrm{exp}}$, as onsets of optical spectra instead of GW band gaps, $E_{\mathrm{g}}^{\mathrm{GW}}$,
because there are differences on the order of 1 eV between GW and experimental band gaps in wide-gap insulators. As shown below,
these differences are comparable to the exciton binding energies in the materials under consideration, and can therefore cause an
artificial cancellation of the two errors in $E_{\mathrm{g}}$ and $E_{\mathrm{b}}$ when one compares the excitonic peak position in the calculated optical spectrum using $E_{\mathrm{g}}^{\mathrm{GW}}$ with the experimental optical spectrum.

In the Dyson equation for optical spectra, we used a 16$\times$16$\times$16 Monkhorst-Pack $\mathbf{k}$-point mesh, 4 valence bands, and 20 conduction bands. We found that TDDFT-LRC shows a slower convergence with respect to the number of conduction bands ($N_{\mathrm{c}}$) than the BSE (e.g. for LiF, $N_{\mathrm{c}}=6$ is enough for the BSE,~\cite{Rohlfing98} while $N_{\mathrm{c}}\ge12$ is needed for both Dyson and Casida equations). An insufficient number of conduction bands causes blueshifts of  the excitonic peak
(i.e. decreases the exciton binding energy) and reduces its oscillator strength in the LRC spectrum of wide-gap insulators significantly. This slow convergence also occurs for the real part of $\epsilon_{\mathrm{M}}^{\mathrm{LRC}}(\omega=0)$ (i.e., the LRC dielectric constant).

In the Dyson equation for Bootstrap-type kernel parameters, we used a 20$\times$20$\times$20 (20$\times$20$\times$10) $\Gamma$-centered $\mathbf{k}$-point mesh, 4 (8) valence bands, 20 (20) conduction bands, and 59 (73) $\mathbf{G}$ vectors for GaAs, $\beta$-GaN, MgO, LiF, solid Ar, and solid Ne ($\alpha$-GaN and AlN).

In the Casida equation, we used a 28$\times$28$\times$28 $\Gamma$-centered $\mathbf{k}$-point mesh, 3 valence bands, 2  conduction bands, and 59
$\mathbf{G}$ vectors for GaAs.
The corresponding parameters for the other materials are: 16$\times$16$\times$16, 3, 6, 59 for $\beta$-GaN and MgO,
16$\times$16$\times$8, 6, 9, 73 for $\alpha$-GaN and AlN, and
8$\times$8$\times$8, 3, 24, 59 for LiF, solid Ar, and solid Ne.

To calculate $\alpha_{\mathrm{JGM}}$, we used a 8$\times$8$\times$8 (8$\times$8$\times$4) $\Gamma$-centered
$\mathbf{k}$-point mesh and 59 (73) $\mathbf{G}$ vectors for GaAs, $\beta$-GaN, MgO, LiF, solid Ar, and solid Ne ($\alpha$-GaN and AlN).

All computational parameters listed here were chosen after performing systematic convergence tests.

\begin{table}[t!]
  \caption{LRC kernel parameters $\alpha$ and exciton binding energies $E_{\mathrm{b}}$ (in meV) obtained from the RPA-Bootstrap kernel using different types of the scissors shift.}
  \begin{tabular*}{0.48\textwidth}{ @{\extracolsep{\fill}} c c c c c c }
    \hline \hline
\multicolumn{2}{c}{Scissors shift} & \multicolumn{2}{c}{GaAs} & \multicolumn{2}{c}{Ne} \\
\cline{1-2} \cline{3-4} \cline{5-6}
$E_{c\mathbf{k}} \to E_{c\mathbf{k}}+\Delta$	& $\hat{p} \to \hat{p}_{\mathrm{renorm}}$	& $\alpha$ 	& $E_{\mathrm{b}}$	& $\alpha$ 	& $E_{\mathrm{b}}$	\\ \hline
Yes											& Yes										& 0.116		& 0.601 				& 37.5 		& 6000 				 \\
Yes											& No										& 0.284		& 0.246 				& 66.7 		& 7420 				 \\
No											& No										& 0.073		& 0.344 				& 30.9 		& 2400 				 \\
    \hline \hline
  \end{tabular*}
  \label{tab:scissors.shift}
\end{table}

\subsection{Effect of the scissors shift on LRC results} \label{subsec:IID}

The exact xc kernel can be written in  the form $f_{\rm xc} = f_{\rm xc}^{\rm qp} + f_{\rm xc}^{\rm ex}$,\cite{Stubner2004,Ullrich2012}
where the ``quasiparticle'' part, $f_{\rm xc}^{\rm qp}$, is responsible for correcting the Kohn-Sham gap, and
$f_{\rm xc}^{\rm ex}$ is the excitonic part. In the standard TDDFT approach for semiconductors and insulators,\cite{Onida2002,Ullrich2015,Sharma2014a} the
quasiparticle part of $f_{\rm xc}$ is ignored, and any corrections to the Kohn-Sham gap are made directly in the input band structure,
usually by means of GW or hybrid functionals; the remaining
part of the xc kernel, $f_{\rm xc}^{\rm ex}$, is then approximated.

A standard and inexpensive method for correcting LDA band structures is by applying the so-called scissors shift.\cite{Levine89,Gonze97}
There are several ways of applying the scissors shift to Dyson and Casida equations in Eqs.~(\ref{eq:chi0}), (\ref{eq:chi0.00}), (\ref{eq:Fxc}), and (\ref{eq:FHxc}) and LRC-type kernels. The scissors shift can be applied to only the conduction bands (i.e. replacing $E_{c\mathbf{k}}$ by $E_{c\mathbf{k}}+\Delta$) or to the momentum operator as well (i.e. replacing $\hat{p}$ by $\hat{p}_{\mathrm{renorm}} = \{(E_{c\mathbf{k}}+\Delta-E_{v\mathbf{k}})/(E_{c\mathbf{k}}-E_{v\mathbf{k}})\}\hat{p}$, where $\hat{p}_{\mathrm{renorm}}$ is the renormalized momentum operator),~\cite{Levine89,DelSole93}
where $\Delta$ is the difference between the experimental (or GW) and DFT band gaps.

Excitonic effects within the LRC approximation are quite sensitive to the particular implementation of the scissors shift.
For example, Table~\ref{tab:scissors.shift} shows $\alpha$ and $E_{\mathrm{b}}$ of GaAs and solid Ne obtained from the RPA-Bootstrap kernel using different types of the scissors shift. We find that the scissors shift affects the LRC results significantly.

In the following, we choose not to apply the scissors shift to $E_{c\mathbf{k}}$ and $\hat{p}$, i.e.,
we calculate exciton binding energies from the Casida equation using the uncorrected LDA band structure as input. Optical spectra, obtained from the Dyson-equation approach, are also calculated based on the uncorrected LDA band structure, and then rigidly shifted to align them
with the experimental band gap. We have chosen this approach for several reasons.

Firstly, the scissors shift is not related to excitons. The scissors shift is a matter of choice for the study of excitonic effects in solids. Our aim is to make the simplest choices (e.g. the LDA and the head-only kernel) and to focus on identifying the origin of conflicting results in existing kernels and designing a new kernel. Note that we applied the scissors shift to $E_{c\mathbf{k}}$ and $\hat{p}$ to reproduce the results of existing kernels, which are provided in the Supplemental Material.\cite{supplemental}

Secondly, we were concerned about the unphysically strong influence of the scissors shift on the LRC results. The scissors shift increases $\epsilon_{00}^{-1}$ by $\sim$10\%; this small increase in $\epsilon_{00}^{-1}$ affects the LRC results for wide-gap insulators significantly when the LRC-type kernel depends on the dielectric constant (see below). In other words, $f_{\mathrm{xc}}^{\mathrm{ex}}=f_{\mathrm{xc}}^{\mathrm{ex}}(f_{\mathrm{xc}}^{\mathrm{qp}})$, but this is not what $f_{\mathrm{xc}}^{\mathrm{qp}}$ and $f_{\mathrm{xc}}^{\mathrm{ex}}$ are meant to be. The big change in the LRC results due to the scissors shift is not associated with excitons.

Thirdly, it allows us to eliminate one source of conflicting results. Some kernels use $E_{\mathrm{g}}^{\mathrm{GW}}$ in the scissors shift, compare their optical spectra with experiment, and reproduce or predict the excitonic peak position for wide-gap insulators by interpreting the $\sim$1~eV error in $E_{\mathrm{g}}^{\mathrm{GW}}$ as $E_{\mathrm{b}}$.~\cite{Sharma11,Berger15} In addition, the small difference between $E_{\mathrm{g}}^{\mathrm{exp}}$ and $E_{\mathrm{g}}^{\mathrm{GW}}$ (or $E_{\mathrm{g}}^{\mathrm{GW}}$ obtained from different potential methods) makes a small difference in the scissors-shifted dielectric constant, which can cause a big difference in the LRC results for wide-gap insulators. By not using the scissors shift, we can avoid these unnecessary complications.

Lastly, by not using the scissors shift we can avoid expensive $E_{\mathrm{g}}^{\mathrm{GW}}$ calculations for unknown materials when we need only $E_{\mathrm{b}}$. When $E_{\mathrm{g}}^{\mathrm{GW}}$ is not calculated and the dielectric constant is calculated from density-functional perturbation theory (DFPT),~\cite{Baroni01,Gajdos06} which is computationally much cheaper than the sum-over-states (SOS) method (\ref{eq:dielectric.function}) because conduction bands are not needed,
large-scale or high-throughput screening exciton calculations become possible.

\begin{figure}[t]
\includegraphics[width=\linewidth]{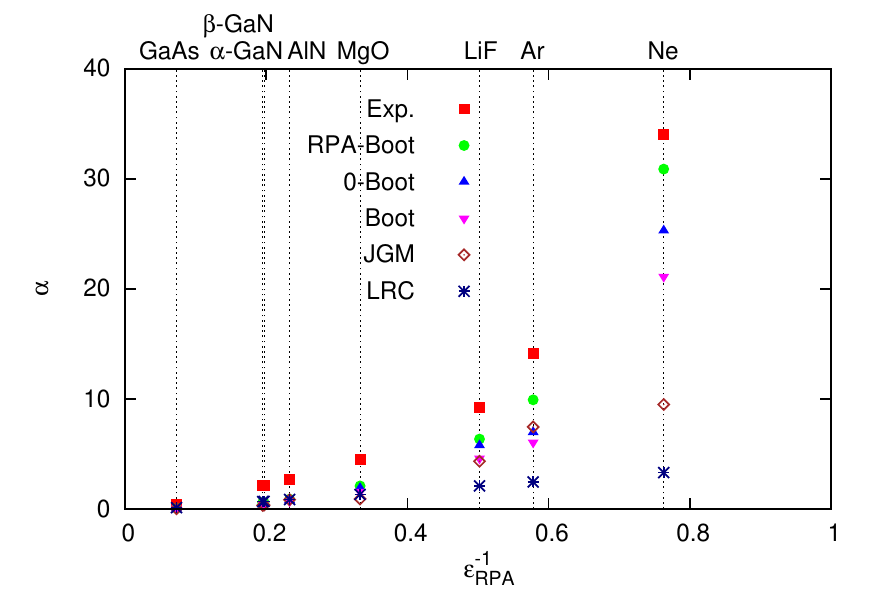}
\caption{(Color online) LRC kernel $\alpha$-parameters for various materials, compared with the $\alpha$-parameter fitted to
reproduce the experimental exciton binding energy (see text).}
\label{fig1}
\end{figure}

%=====================================================================================================================================
\section{Results and Discussion} \label{sec:III}

\subsection{Comparison of LRC $\alpha$-parameters} \label{subsec:IIIA}

In the following, we will discuss our results for the excitonic properties of the bulk semiconductors GaAs, $\alpha$-GaN, and $\beta$-GaN,
the narrow-gap insulators AlN and MgO, and the wide-gap insulators LiF, solid Ar, and solid Ne.
The experimental exciton binding energies are obtained from Refs. \onlinecite{PCK92,ASWS97,MLSK97,HKKS69,RW67,SK79,Leute09,Roessler67}.
We point out again that all results shown below were obtained without using the scissors shift.

Let us begin with an assessment of the LRC $\alpha$-parameters for various materials. Figure~\ref{fig1}
compares $\alpha_{\rm LRC}$, $\alpha_{\rm JGM}$, $\alpha_{\rm B}$, $\alpha_{\rm 0B}$, and $\alpha_{\rm RPAB}$
with the $\alpha$-parameter $\alpha_{\mathrm{exp}}$ which, when used in the head-only LRC kernel (\ref{LRCh}), reproduces the experimental
exciton binding energy for each material under consideration.
We see that $\alpha$ varies from $\sim$0.1 ($\alpha_{\mathrm{RPAB}}$ for GaAs) to $\sim$30 ($\alpha_{\mathrm{RPAB}}$ for solid Ne).
All calculated $\alpha$-parameters are smaller than the experimentally fitted ones.

\begin{figure}[t]
\begin{tabular}{c}
{\includegraphics[trim=0mm 0mm 0mm 0mm, clip, width=0.48\textwidth]{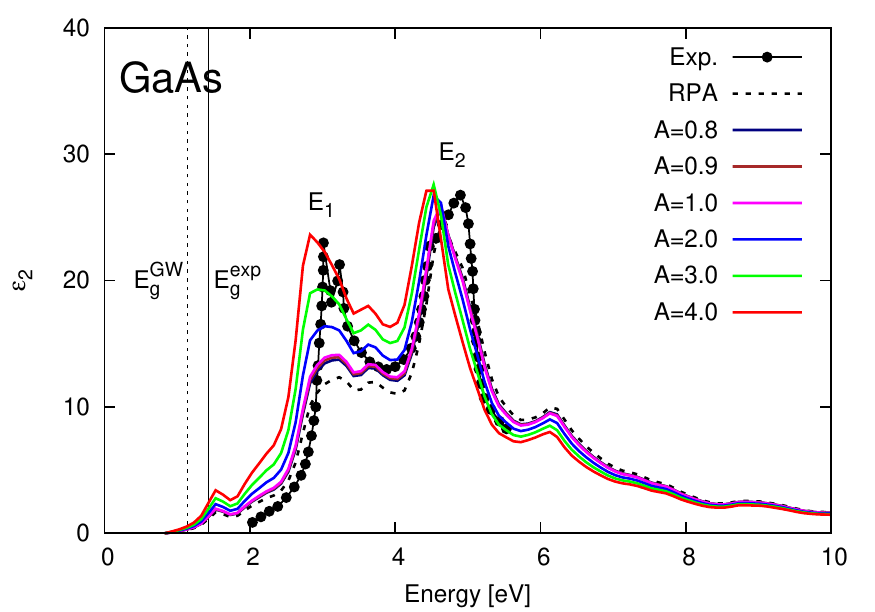}} \\
{\includegraphics[trim=0mm 0mm 0mm 0mm, clip, width=0.48\textwidth]{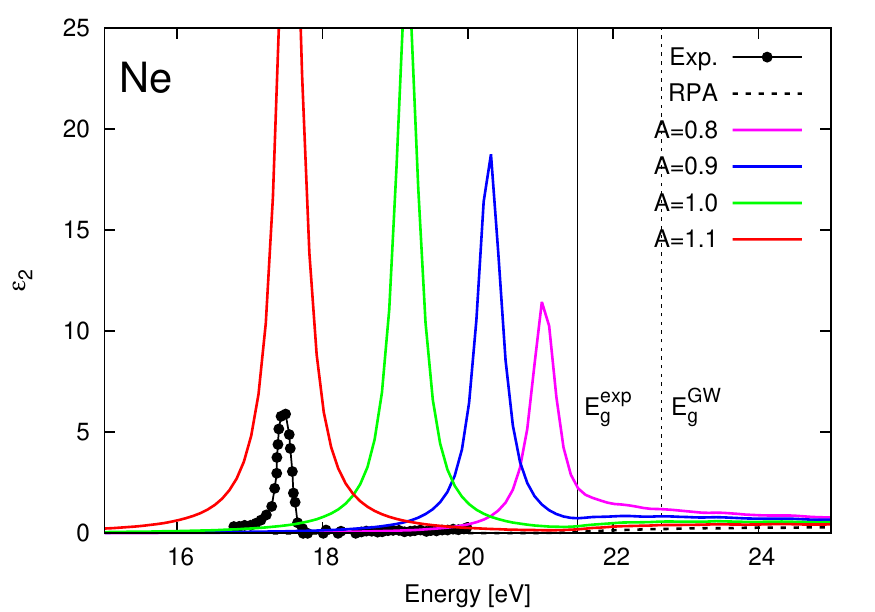}}
\end{tabular}
\caption{(Color online) Experimental and calculated optical absorption spectra of GaAs (top) and solid Ne (bottom). For the LRC kernel, $\alpha=A\alpha_{\mathrm{RPAB}}$ is used, where $\alpha_{\mathrm{RPAB}} = 0.073$ and 30.9 for GaAs and solid Ne, respectively. Note that $A = 0.8$ and 0.9 approximately correspond to Bootstrap and 0-Bootstrap kernels, respectively. The $E_{\mathrm{1}}$ and $E_{\mathrm{2}}$ peaks in the spectrum of GaAs are at critical points, where conduction and valence bands are parallel to each other. }
\label{fig2}
\end{figure}

\subsection{Sensitivity of optical spectra to changes in $\alpha$} \label{subsec:IIIB}

Next, we examine the effects of the head-only LRC kernel on optical absorption spectra. Figure~\ref{fig2} shows calculated
optical spectra of GaAs and solid Ne obtained from the Dyson equation using the LRC kernel with $\alpha=A\alpha_{\mathrm{RPAB}}$,
where $A$ is a scaling factor, and compares them with experiment.\cite{PCK92,SK79}
We chose GaAs and solid Ne because they are
extreme examples of semiconductors with weakly bound Wannier-Mott excitons and insulators with strongly bound Frenkel excitons.
In the case of GaAs, the optical spectrum shows two prominent peaks above the band gap; $E_1$ can be interpreted as a continuum exciton.
The bound exciton below the gap is very weak, and not visible on the scale of this plot because  $E_{\rm b}$ is much smaller than the line broadening. To see the bound Wannier-Mott exciton of GaAs, high-resolution spectroscopy at low temperatures is needed.\cite{Ulbrich1985}
On the other hand, for solid Ne the excitonic peak
is very prominent and far from the gap, and it is easy to obtain $E_{\rm b}$ from the spectrum.

In the top panel of Fig.~\ref{fig2}, we show calculated optical spectra of GaAs for a range of $A$ between 0.8 and 4.0. We find that
the spectra are rather insensitive to the scaling: a 10\% change in $\alpha$ has only a very small effect: in other words, $\alpha$ has a big margin for semiconductors. The RPA spectrum
of GaAs is already quite similar to experiment, apart from the height of the $E_1$-peak. To obtain the experimental height
of the $E_1$-peak, a scaling factor of $A\approx4$ (i.e. $\alpha\approx0.3$) is needed; however, this also increases the peak width, and the valley between the $E_1$ and the $E_2$ peak becomes too high.

The bottom panel of Fig.~\ref{fig2} shows the calculated spectra of solid Ne for a much smaller range of $A$, between 0.8 and 1.1.
Here, the spectra are very sensitive to the change in $\alpha$: a 10\% change shifts excitonic peaks by about 1 eV: in other words,
$\alpha$ has a small margin for insulators.  Clearly, the RPA spectrum of Ne is completely different from experiment, and the LRC kernel reshapes it significantly. Using $A\approx1.1$ puts the excitonic peak
at the right position; however, the peak height and width is now drastically overestimated.

The low sensitivity of LRC results for semiconductors to changes in $\alpha$ explains why there are so many LRC-type kernels. LRC-type kernels only slightly modify RPA spectra of semiconductors, which are already very close to experiment, and $\alpha$ has a big margin for semiconductors. Thus, all LRC-type kernels produce similar and seemingly good optical spectra of semiconductors even when they use different choices and yield very different $\alpha$ values (e.g. $\alpha_{\mathrm{LRC}}\approx0.2$ and $\alpha_{\mathrm{B}}\approx0.1$ for Si and GaAs).

The high sensitivity of LRC results for insulators to changes in $\alpha$ is consistent with the finding of Ref.~\onlinecite{Rigamonti15}. The idea of the RPA-Bootstrap kernel is to increase $E_{\mathrm{b}}$ for wide-gap insulators from $\sim$0.1~eV to $\sim$1~eV by increasing $\alpha_{\mathrm{B}}$ by $\sim$20\% for all materials. However, the $\sim$20\% increase in $\alpha_{\mathrm{B}}$ does not fix the problem of the Bootstrap kernel of not reproducing an excitonic peak in the optical spectrum of semiconductors such as Si, because of the low
$\alpha$-sensitivity of LRC results for semiconductors.

We also point out that the LRC results show a similar sensitivity trend to diagonal versus head-only LRC kernels and full versus TDA Casida equations (i.e. insensitive for semiconductors, but sensitive for insulators).~\cite{Byun2017}

These two examples already indicate a general limitation of the LRC kernel that applies to \emph{all} materials: it is impossible to obtain the correct position and the correct height and width of an excitonic peak in the LRC spectrum, for \emph{both} semiconductors and insulators. We will provide more evidence for this conclusion and give more examples below.
To reproduce a given excitonic feature for both semiconductors and insulators (e.g., the peak height or the peak position), it is clear that a nonuniform scaling factor for Bootstrap-type kernels will be needed: the scaling factor should be close to 1 for insulators, but much greater than 1 for semiconductors. Any method which nearly uniformly scales Bootstrap-type kernels for all materials [such as using different dielectric constants (e.g. bootstrapped vs not, scissors-shifted vs not, macroscopic vs microscopic, or RPA vs LDA, all of which are different from each other by $\sim$10\%) in the numerator and/or the denominator of Eqs.~(\ref{eq:fxc.fB}), (\ref{eq:fxc.hB}), (\ref{eq:fxc.0B}), and (\ref{eq:fxc.RPAB}) or using different band structures] is likely to fail to produce satisfactory results across the board.

\begin{figure}[t]
\includegraphics[trim=0mm 0mm 0mm 0mm, clip, width=0.48\textwidth]{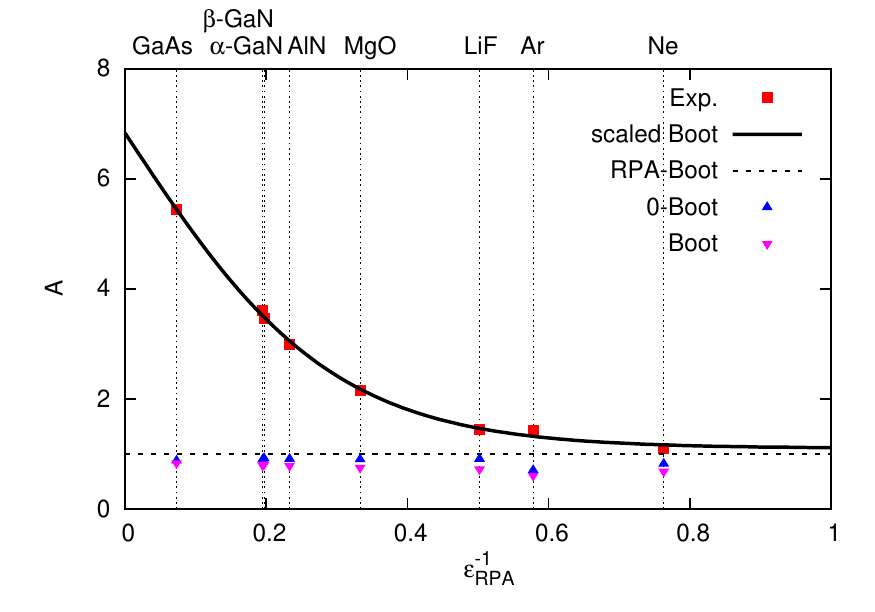}
\caption{(Color online) Scaling factors $A_{\rm B}$, $A_{\rm 0B}$ and $A_{\rm exp}$ (see text) of Bootstrap-type kernels for various materials. }
\label{fig3}
\end{figure}

\begin{figure}[t]
\includegraphics[trim=0mm 0mm 0mm 0mm, clip, width=0.48\textwidth]{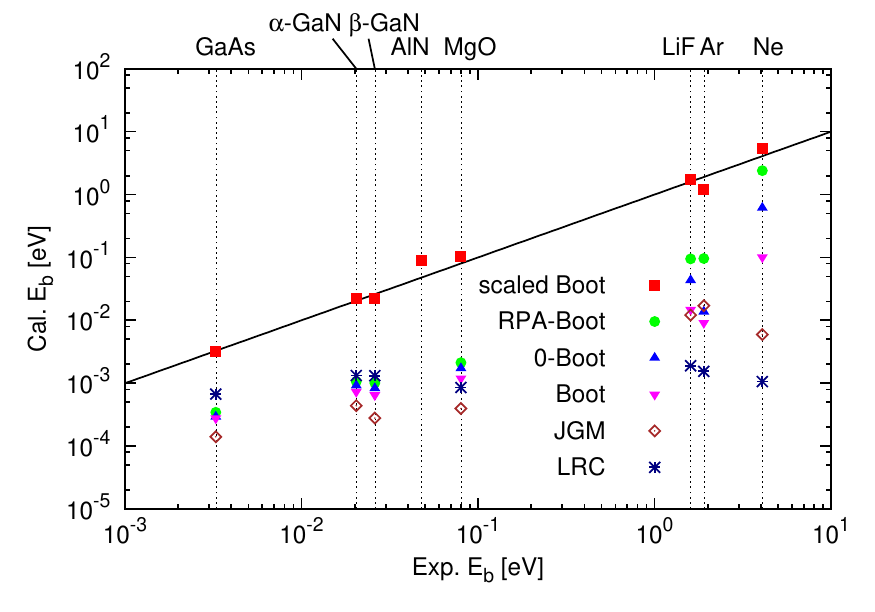}
\caption{(Color online) Experimental and calculated exciton binding energies $E_{\mathrm{b}}$. }
\label{fig4}
\end{figure}

\begin{table*}
  \caption{Experimental and calculated exciton binding energies $E_{\mathrm{b}}$ (in meV).}
%  \begin{tabular}{ c c c c c c c c c c c c }
  \begin{tabular*}{0.96\textwidth}{ @{\extracolsep{\fill}} r c c r r r r r r r r r r }
    \hline \hline
				&										&						&	GaAs &	$\alpha$-GaN &	$\beta$-GaN &	AlN	&	MgO &	LiF 	&	 Ar 	&	 Ne    	\\ \hline
Exp.				&										&						&	3.27 &			20.4 &			26.0 &	48.0 &	80.0 &	1600 &	 1900 &	 4080 	\\ \hline
scaled Boot		& $\epsilon_{\mathrm{LDA(DFPT)}}^{-1}$	& Eq.~(\ref{eq:A.FDd})	&	3.30 &			23.1 &			21.4 &	97.4 &	90.1 &	1790 &	 1230 &	5220 	 \\
scaled Boot		& $\epsilon_{\mathrm{LDA(DFPT)}}^{-1}$	& Eq.~(\ref{eq:A.exp})	&	3.30 &			23.0 &			21.4 &	97.4 &	92.3 &	1790 &	 1230 &	5190 	 \\
scaled Boot		& $\epsilon_{\mathrm{RPA(SOS)}}^{-1}$	& Eq.~(\ref{eq:A.FDd})	&	3.24 &			22.2 &			22.1 &	90.4 &	97.2 &	1710 &	 1220 &	5410 	 \\
scaled Boot		& $\epsilon_{\mathrm{RPA(SOS)}}^{-1}$	& Eq.~(\ref{eq:A.exp})	&	3.24 &			22.1 &			22.0 &	90.6 &	102 &	1720 &	 1210 &	5350 	 \\ \hline
RPA-Boot		&										&						&	0.344 &			1.06 &			1.01 &	0.00 &	2.12 &	94.7 &	 96.0 &	2400 	 \\
0-Boot			&										&						&	0.293 &			0.919 &			0.829 &	0.00 &	1.72 &	43.2 &	 13.7 &	612 	 \\
Boot			&										&						&	0.278 &			0.735 &			0.649 &	0.00 &	1.20 &	14.8 &	 9.14 &	101 	 \\
JGM				&										&						&	0.141 &			0.438 &			0.279 &	0.00 &	0.397 &	12.1 &	 17.1 &	5.96 	 \\
LRC				&										&						&	0.670 &			1.33 &			1.32 &	0.00 &	0.855 &	1.89 &	 1.54 &	1.06 	 \\
    \hline \hline
  \end{tabular*}
%  \end{tabular}
  \label{tab:Eb}
\end{table*}

\subsection{Nonuniformly scaled Bootstrap kernel} \label{subsec:IIIC}

In Fig.~\ref{fig1} we compared the $\alpha$ values from head-only LRC-type kernels for various materials,
and found that for wide-gap insulators, $\alpha_{\mathrm{RPAB}}$ shows the most similar trend to
$\alpha_{\mathrm{exp}}$ (e.g. Bootstrap and 0-Bootstrap kernels yield $E_{\mathrm{b}}$ of solid Ar that is smaller than that of LiF). We therefore choose it as the basis for constructing a new, scaled Bootstrap xc kernel.

%(especially, solid Ar)

Let us first define $f_{\rm xc}^{\rm B(h)} = A_{\rm B} f_{\rm xc}^{\rm RPAB}$ and
$f_{\rm xc}^{\rm 0B} = A_{\rm 0B} f_{\rm xc}^{\rm RPAB}$. The values of $A_{\rm B}$ and $A_{\rm 0B}$ are plotted
in Fig.~\ref{fig3} as a function of $\epsilon_{\rm RPA}^{-1}$ for various materials; we find that $A_{\mathrm{B}}\approx0.8$ and $A_{\mathrm{0B}}\approx0.9$ for all materials (i.e. Bootstrap-type kernels are nearly uniformly scaled to each other). On the other hand,
if we define $f_{\rm xc}^{\rm exp} = A_{\rm exp} f_{\rm xc}^{\rm RPAB}$ (i.e. $\alpha_{\mathrm{exp}}=A_{\mathrm{exp}}\alpha_{\mathrm{RPAB}}$) as the head-only LRC xc kernel which
reproduces the experimental exciton binding energy, we can see that $A_{\rm exp}$ varies strongly as a function
of material, from $\sim$1.1 (solid Ne) to $\sim$5 (GaAs).
This non-uniform variation is consistent with our observations from the optical spectra of GaAs and solid Ne, see Fig. \ref{fig2}.

The values of $A_{\rm exp}$ show a rather smooth behavior as a function of $\epsilon_{\rm RPA}^{-1}$, which suggests
that a scaled Bootstrap kernel can be defined via a fit to the experimental data:
\begin{equation}
f_{\mathrm{xc}}^{\rm sB} = A(x) f_{\mathrm{xc}}^{\mathrm{RPAB}} = - A(x) \frac{4\pi x}{(1/x - 1) \mathbf{q}^2}, \label{eq:NUS.Bootstrap}
\end{equation}
where $x=\epsilon_{\mathrm{RPA}}^{-1}$ (alternatively, choosing $x=\epsilon_{\mathrm{LDA}}^{-1}$ would have been possible as well).
Note that both SOS and DFPT methods yield the same $x$ value.~\cite{Gajdos06} Among many $\epsilon^{-1}$, we used $\epsilon_{\mathrm{RPA(SOS)}}^{-1}$ and $\epsilon_{\mathrm{LDA(DFPT)}}^{-1}$ in this work because they can be
easily obtained from the Abinit code. Among the two $\epsilon^{-1}$, we used $\epsilon_{\mathrm{RPA(SOS)}}^{-1}$ to obtain $E_{\mathrm{b}}$ and optical spectra in this work unless stated otherwise.

We found two fitting functions, which describe well the non-uniformity of $A_{\rm exp}$,
\begin{align}
A(x) &= \frac{a_{1}}{e^{(x-a_{2})/a_{3}} + 1} + a_{4}, \label{eq:A.FDd} \\
   &= b_{1}e^{-x^{b_{2}}/b_{3}} + b_{4}. \label{eq:A.exp}
\end{align}
The fitting parameters $a_i$ and $b_i$, where $i = 1, 2, 3, 4$, are summarized in Table~\ref{tab:a.b}. Note that these fitting parameters
are appropriate for the specific choices made here: experimental lattice constant, pseudopotential method, LDA band structure, head-only LRC kernel, LFE, and no scissors shift. If other choices are made, such as an all-electron method or using the scissors shift, one needs to re-fit the parameters $a_i$ and $b_i$. This calibration is inevitable due to the high sensitivity of the LRC results for wide-gap insulators.
We found that the two fitting functions yield almost the same result for $A$ (and thus $\alpha$ and $E_{\mathrm{b}}$) except for $\epsilon_{\infty} \gg 10$ (see Table~\ref{tab:Eb}). Among the two fitting functions, we chose to use Eq.~(\ref{eq:A.exp}) to obtain $E_{\mathrm{b}}$ and optical spectra.

\begin{table}
  \caption{Fitting parameters for the scaling factor $A$.}
  \begin{tabular*}{0.48\textwidth}{ @{\extracolsep{\fill}} c | c c c c | c c c c }
    \hline \hline
$x$										& $a_1$	& $a_2$		& $a_3$	& $a_4$	& $b_1$	& $b_2$	& $b_3$	& $b_4$	\\ \hline
$\epsilon_{\mathrm{LDA(DFPT)}}^{-1}$	& 25.9	& -0.159		& 0.161	& 1.16	& 6.89	& 1.11	& 0.166	& 1.16	\\
$\epsilon_{\mathrm{RPA(SOS)}}^{-1}$	& 11.6	& -0.00239	& 0.148	& 1.10	& 5.56	& 1.25	& 0.155	& 1.11	\\
    \hline \hline
  \end{tabular*}
  \label{tab:a.b}
\end{table}

Figure~\ref{fig4} and Table~\ref{tab:Eb} show experimental and calculated $E_{\mathrm{b}}$ of various materials. Whereas other kernels underestimate $E_{\mathrm{b}}$ for all materials by $\sim$10 times, the scaled Bootstrap kernel yields accurate and consistent $E_{\mathrm{b}}$. The most significant
deviations are for AlN (where all other kernels give zero exciton binding energy) and for solid Ar (where even the
BSE underestimates $E_{\rm b}$  by $\sim$0.3~eV~\cite{Sottile07}).

\begin{figure*}
\begin{tabular}{c c}
{\includegraphics[trim=0mm 0mm 0mm 0mm, clip, width=0.48\textwidth]{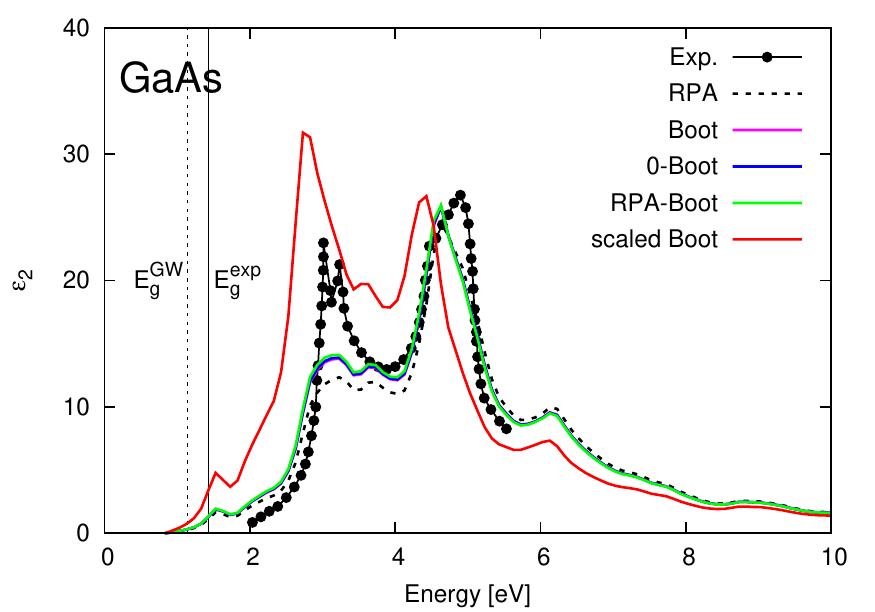}}
{\includegraphics[trim=0mm 0mm 0mm 0mm, clip, width=0.48\textwidth]{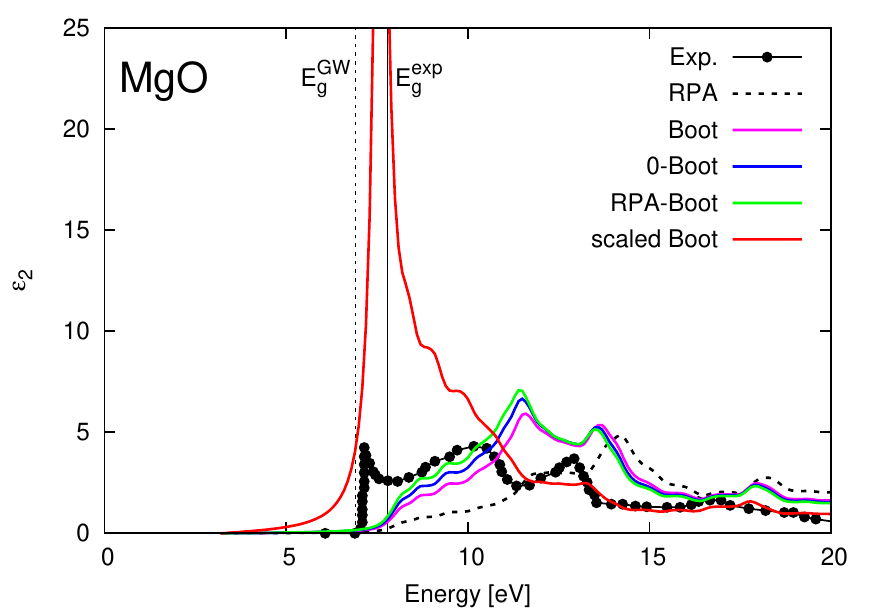}} \\
{\includegraphics[trim=0mm 0mm 0mm 0mm, clip, width=0.48\textwidth]{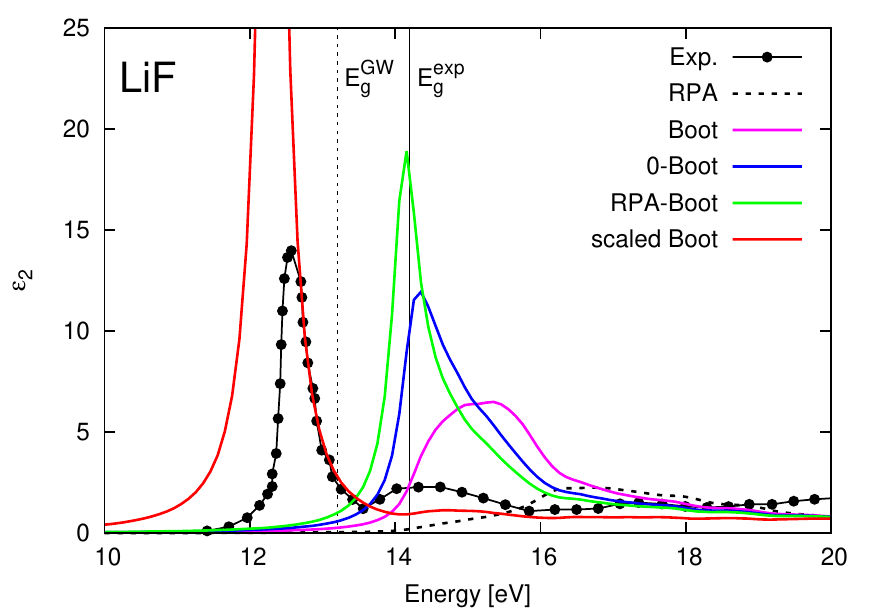}}
{\includegraphics[trim=0mm 0mm 0mm 0mm, clip, width=0.48\textwidth]{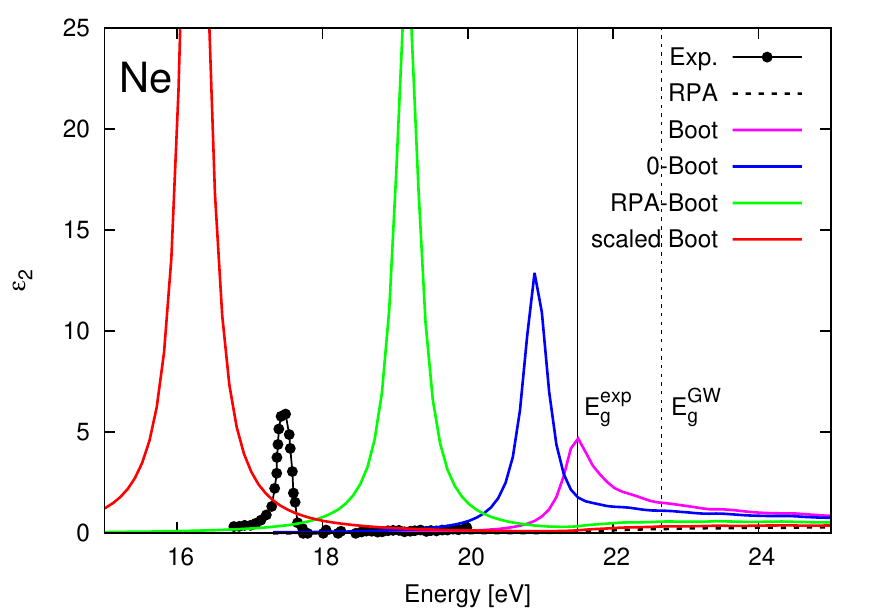}}
\end{tabular}
\caption{(Color online) Experimental and calculated optical absorption spectra of GaAs, MgO, LiF, and solid Ne. }
\label{fig5}
\end{figure*}

Figure~\ref{fig5} shows experimental and calculated optical spectra of GaAs, MgO, LiF, and solid Ne.
We included LiF because it is one of two extreme examples of wide-gap insulators.
We also included MgO because the LRC spectrum of MgO is very different from experiment at all $\alpha$ values, so it is impossible to determine an optimal $\alpha$ value for MgO by varing $\alpha$ (i.e., no $\alpha$ exists that reproduces the correct excitonic peak height or position).~\cite{Botti04} Here, we report the LRC spectrum of MgO when $\alpha \approx \alpha_{\mathrm{exp}}$.
Bootstrap-type kernels with similar $\alpha$ values produce very similar optical spectra of GaAs (a semiconductor) and MgO (a narrow-gap insulator), but very different ones of LiF and solid Ne (wide-gap insulators). As discussed earlier, this is due to the different sensitivity of LRC spectra to semiconductors and insulators.

Our scaled Bootstrap kernel, which is designed to reproduce $E_{\mathrm{b}}^{\mathrm{exp}}$, yields excitonic peaks with overestimated (i.e. higher and wider) oscillator strengths in optical spectra of GaAs and MgO, while other Bootstrap-type kernels, which underestimate $E_{\mathrm{b}}$ by $\sim$10 times, barely produce excitonic peaks. This indicates that the LRC kernel cannot produce correct exciton binding energies and optical spectra at the same time for \emph{all} materials (i.e. for semiconductors {\em and} insulators). Our finding is consistent with the LRC spectrum of ZnO, in which the calculated excitonic peak is much higher and wider than the experimental one.\cite{Gori10}

We emphasize that our kernel is empirical, but \emph{universal} in that it works for all materials and all choices. In contrast, the RPA-Bootstrap kernel, one of parameter-free kernels, works only for wide-gap insulators under special conditions such as experimental lattice parameters, the head-only kernel, and the scissors shift. In principle, a parameter-free LRC-type kernel cannot be universal for all choices due to the high sensitivity of LRC results for wide-gap insulators; thus, a trade-off between parameter-free and universal is unavoidable.

Our scaling approach is not just another Bootstrap-type kernel or a correction to the RPA-Bootstrap kernel: it is a method to predict $E_{\mathrm{b}}$ of unknown materials using the experimental $E_{\mathrm{b}}$  of a few known materials as input. The RPA-Bootstrap kernel is merely used as a fitting function, which was chosen to demonstrate the problems of popular Bootstrap-type kernels \cite{Sharma11,Rigamonti15,Berger15} and to suggest a simple way to fix them. One has the full freedom to use any other LRC-type fitting functions for our method.

\section{Conclusions} \label{sec:IV}

In this paper, we have carried out a systematic numerical assessment of the family of
static long-range-corrected (LRC) xc kernels for
solids. The main challenge faced by TDDFT for the optical spectral properties of semiconductors and insulators
is to reproduce the excitonic peaks at the right position and with the correct strength. We have used two methods: the Dyson-equation
approach, which yields optical spectra, and the Casida-equation approach, which allows a precise determination of exciton binding
energies. The two methods are equivalent, i.e., they give, in principle, the same excitonic peak positions, but in their practical implementations
they are very different: from the Dyson equation approach, and the resulting macroscopic dielectric function, one cannot extract
the binding energies of weakly bound excitons. Hence, the Casida approach is a very useful method, complementing the standard Dyson
approach.

We have studied a group of materials, ranging from small-gap semiconductors to large-gap insulators, with exciton binding energies
between a few meV and several eV. For these materials, we have tested the empirical LRC kernel, several flavors of the Bootstrap kernel,
and the jellium-with-a-gap model. Most of these methods produce decently-looking optical spectra for semiconductors, but the
exciton binding energies are consistently underestimated. We proposed a new xc kernel, obtained via a material-dependent scaling
of the RPA-Bootstrap kernel. The scaled Bootstrap kernel is designed to produce accurate exciton binding energies for all materials
under study, at very low computational cost. However, there is a price to pay: it turns out that it is impossible to obtain accurate
exciton binding energies and good optical spectra at the same time for all materials using any LRC method---if the exciton peak is at the right place,
the oscillator strength (i.e., the peak height and width) tends to be exaggerated for both semiconductors and insulators.

In general, assessing the performance of xc kernels for excitonic properties is a delicate task, because there are many choices involved.
Here, we chose to use LDA band structures obtained with a pseudopotential code, we included local-field effects, and we implemented
the xc kernels in their head-only forms. These choices will affect the numerical results: whereas the spectra of semiconductors are
relatively insensitive to the strength $\alpha$ of the head of the LRC kernel, the spectra of insulators are very sensitive. Hence,
it is crucial that all choices made are clearly identified, in order to facilitate comparison between results obtained by different
research groups.

The main outcome of our work is that we have developed a method which can produce accurate exciton binding energies at a low
computational cost. In practice, the parameters for the scaling function should be re-fitted for each particular implementation,
using a small test set of small- and large-gap materials. It should then be possible to obtain accurate exciton binding energies
for other, more complicated materials. Such calculations are currently in progress.

The ultimate goal is to develop TDDFT approaches that yield both accurate exciton binding energies and spectral shapes.
As we have seen, the LRC method is too restricted to achieve both. TDDFT is in principle exact; however, going beyond the LRC
approach is very challenging: we may need to better understand the role of the wings and body of the xc kernel,
and the frequency dependence of the xc kernel may have to be taken into account.
Alternatives beyond pure TDDFT, such as
hybrid functionals, are therefore very promising. Such methods are currently under development.

%$f_{\mathrm{xc}}$

\begin{acknowledgments}
This work was supported by NSF grant No. DMR-1408904.
The computation for this work was performed on the high performance computing infrastructure provided by Research Computing Support Services at the University of Missouri-Columbia. We thank Lucia Reining and Sangeeta Sharma for discussions and for providing valuable comments
on the manuscript.
\end{acknowledgments}

\bibliography{Bootstrap_refs}

\begin{thebibliography}{54}%
\makeatletter
\providecommand \@ifxundefined [1]{%
 \@ifx{#1\undefined}
}%
\providecommand \@ifnum [1]{%
 \ifnum #1\expandafter \@firstoftwo
 \else \expandafter \@secondoftwo
 \fi
}%
\providecommand \@ifx [1]{%
 \ifx #1\expandafter \@firstoftwo
 \else \expandafter \@secondoftwo
 \fi
}%
\providecommand \natexlab [1]{#1}%
\providecommand \enquote  [1]{``#1''}%
\providecommand \bibnamefont  [1]{#1}%
\providecommand \bibfnamefont [1]{#1}%
\providecommand \citenamefont [1]{#1}%
\providecommand \href@noop [0]{\@secondoftwo}%
\providecommand \href [0]{\begingroup \@sanitize@url \@href}%
\providecommand \@href[1]{\@@startlink{#1}\@@href}%
\providecommand \@@href[1]{\endgroup#1\@@endlink}%
\providecommand \@sanitize@url [0]{\catcode `\\12\catcode `\$12\catcode
  `\&12\catcode `\#12\catcode `\^12\catcode `\_12\catcode `\%12\relax}%
\providecommand \@@startlink[1]{}%
\providecommand \@@endlink[0]{}%
\providecommand \url  [0]{\begingroup\@sanitize@url \@url }%
\providecommand \@url [1]{\endgroup\@href {#1}{\urlprefix }}%
\providecommand \urlprefix  [0]{URL }%
\providecommand \Eprint [0]{\href }%
\providecommand \doibase [0]{http://dx.doi.org/}%
\providecommand \selectlanguage [0]{\@gobble}%
\providecommand \bibinfo  [0]{\@secondoftwo}%
\providecommand \bibfield  [0]{\@secondoftwo}%
\providecommand \translation [1]{[#1]}%
\providecommand \BibitemOpen [0]{}%
\providecommand \bibitemStop [0]{}%
\providecommand \bibitemNoStop [0]{.\EOS\space}%
\providecommand \EOS [0]{\spacefactor3000\relax}%
\providecommand \BibitemShut  [1]{\csname bibitem#1\endcsname}%
\let\auto@bib@innerbib\@empty
%</preamble>
\bibitem [{\citenamefont {Miyata}\ \emph {et~al.}(2015)\citenamefont {Miyata},
  \citenamefont {Mitioglu}, \citenamefont {Plochocka}, \citenamefont
  {Portugall}, \citenamefont {Wang}, \citenamefont {Stranks}, \citenamefont
  {Snaith},\ and\ \citenamefont {Nicholas}}]{Miyata2015}%
  \BibitemOpen
  \bibfield  {author} {\bibinfo {author} {\bibfnamefont {A.}~\bibnamefont
  {Miyata}}, \bibinfo {author} {\bibfnamefont {A.}~\bibnamefont {Mitioglu}},
  \bibinfo {author} {\bibfnamefont {P.}~\bibnamefont {Plochocka}}, \bibinfo
  {author} {\bibfnamefont {O.}~\bibnamefont {Portugall}}, \bibinfo {author}
  {\bibfnamefont {J.~T.-W.}\ \bibnamefont {Wang}}, \bibinfo {author}
  {\bibfnamefont {S.~D.}\ \bibnamefont {Stranks}}, \bibinfo {author}
  {\bibfnamefont {H.~J.}\ \bibnamefont {Snaith}}, \ and\ \bibinfo {author}
  {\bibfnamefont {R.~J.}\ \bibnamefont {Nicholas}},\ }\href@noop {} {\bibfield
  {journal} {\bibinfo  {journal} {Nature Phys.}\ }\textbf {\bibinfo {volume}
  {11}},\ \bibinfo {pages} {582} (\bibinfo {year} {2015})}\BibitemShut
  {NoStop}%
\bibitem [{\citenamefont {Rohlfing}\ and\ \citenamefont
  {Louie}(1998)}]{Rohlfing98}%
  \BibitemOpen
  \bibfield  {author} {\bibinfo {author} {\bibfnamefont {M.}~\bibnamefont
  {Rohlfing}}\ and\ \bibinfo {author} {\bibfnamefont {S.~G.}\ \bibnamefont
  {Louie}},\ }\href@noop {} {\bibfield  {journal} {\bibinfo  {journal} {Phys.
  Rev. Lett.}\ }\textbf {\bibinfo {volume} {81}},\ \bibinfo {pages} {2312}
  (\bibinfo {year} {1998})}\BibitemShut {NoStop}%
\bibitem [{\citenamefont {Onida}\ \emph {et~al.}(2002)\citenamefont {Onida},
  \citenamefont {Reining},\ and\ \citenamefont {Rubio}}]{Onida2002}%
  \BibitemOpen
  \bibfield  {author} {\bibinfo {author} {\bibfnamefont {G.}~\bibnamefont
  {Onida}}, \bibinfo {author} {\bibfnamefont {L.}~\bibnamefont {Reining}}, \
  and\ \bibinfo {author} {\bibfnamefont {A.}~\bibnamefont {Rubio}},\
  }\href@noop {} {\bibfield  {journal} {\bibinfo  {journal} {Rev. Mod. Phys.}\
  }\textbf {\bibinfo {volume} {74}},\ \bibinfo {pages} {601} (\bibinfo {year}
  {2002})}\BibitemShut {NoStop}%
\bibitem [{\citenamefont {Martin}\ \emph {et~al.}(2016)\citenamefont {Martin},
  \citenamefont {Reining},\ and\ \citenamefont {Ceperley}}]{Reining2016}%
  \BibitemOpen
  \bibfield  {author} {\bibinfo {author} {\bibfnamefont {R.~M.}\ \bibnamefont
  {Martin}}, \bibinfo {author} {\bibfnamefont {L.}~\bibnamefont {Reining}}, \
  and\ \bibinfo {author} {\bibfnamefont {D.~M.}\ \bibnamefont {Ceperley}},\
  }\href@noop {} {\emph {\bibinfo {title} {Interacting Electrons: Theory and
  Computational Approaches}}}\ (\bibinfo  {publisher} {Cambridge University
  Press},\ \bibinfo {address} {Cambridge},\ \bibinfo {year} {2016})\BibitemShut
  {NoStop}%
\bibitem [{\citenamefont {Ullrich}(2012)}]{Ullrich2012}%
  \BibitemOpen
  \bibfield  {author} {\bibinfo {author} {\bibfnamefont {C.~A.}\ \bibnamefont
  {Ullrich}},\ }\href@noop {} {\emph {\bibinfo {title} {Time-dependent
  density-functional theory: concepts and applications}}}\ (\bibinfo
  {publisher} {Oxford University Press},\ \bibinfo {address} {Oxford},\
  \bibinfo {year} {2012})\BibitemShut {NoStop}%
\bibitem [{\citenamefont {Ullrich}\ and\ \citenamefont
  {Yang}(2015)}]{Ullrich2015}%
  \BibitemOpen
  \bibfield  {author} {\bibinfo {author} {\bibfnamefont {C.~A.}\ \bibnamefont
  {Ullrich}}\ and\ \bibinfo {author} {\bibfnamefont {Z.-H.}\ \bibnamefont
  {Yang}},\ }in\ \href@noop {} {\emph {\bibinfo {booktitle} {Density-Functional
  Methods for Excited States}}},\ \bibinfo {series} {Topics in Current
  Chemistry}, Vol.\ \bibinfo {volume} {368},\ \bibinfo {editor} {edited by\
  \bibinfo {editor} {\bibfnamefont {N.}~\bibnamefont {Ferr\'e}}, \bibinfo
  {editor} {\bibfnamefont {M.}~\bibnamefont {Filatov}}, \ and\ \bibinfo
  {editor} {\bibfnamefont {M.}~\bibnamefont {Huix-Rotllant}}}\ (\bibinfo
  {publisher} {Springer},\ \bibinfo {address} {Berlin},\ \bibinfo {year}
  {2015})\ p.\ \bibinfo {pages} {185}\BibitemShut {NoStop}%
\bibitem [{\citenamefont {Reining}\ \emph {et~al.}(2002)\citenamefont
  {Reining}, \citenamefont {Olevano}, \citenamefont {Rubio},\ and\
  \citenamefont {Onida}}]{Reining02}%
  \BibitemOpen
  \bibfield  {author} {\bibinfo {author} {\bibfnamefont {L.}~\bibnamefont
  {Reining}}, \bibinfo {author} {\bibfnamefont {V.}~\bibnamefont {Olevano}},
  \bibinfo {author} {\bibfnamefont {A.}~\bibnamefont {Rubio}}, \ and\ \bibinfo
  {author} {\bibfnamefont {G.}~\bibnamefont {Onida}},\ }\href@noop {}
  {\bibfield  {journal} {\bibinfo  {journal} {Phys. Rev. Lett.}\ }\textbf
  {\bibinfo {volume} {88}},\ \bibinfo {pages} {066404} (\bibinfo {year}
  {2002})}\BibitemShut {NoStop}%
\bibitem [{\citenamefont {Sottile}\ \emph {et~al.}(2003)\citenamefont
  {Sottile}, \citenamefont {Olevano},\ and\ \citenamefont {Reining}}]{SOR03}%
  \BibitemOpen
  \bibfield  {author} {\bibinfo {author} {\bibfnamefont {F.}~\bibnamefont
  {Sottile}}, \bibinfo {author} {\bibfnamefont {V.}~\bibnamefont {Olevano}}, \
  and\ \bibinfo {author} {\bibfnamefont {L.}~\bibnamefont {Reining}},\
  }\href@noop {} {\bibfield  {journal} {\bibinfo  {journal} {Phys. Rev. Lett.}\
  }\textbf {\bibinfo {volume} {91}},\ \bibinfo {pages} {056402} (\bibinfo
  {year} {2003})}\BibitemShut {NoStop}%
\bibitem [{\citenamefont {Adragna}\ \emph {et~al.}(2003)\citenamefont
  {Adragna}, \citenamefont {Del~Sole},\ and\ \citenamefont {Marini}}]{ADM03}%
  \BibitemOpen
  \bibfield  {author} {\bibinfo {author} {\bibfnamefont {G.}~\bibnamefont
  {Adragna}}, \bibinfo {author} {\bibfnamefont {R.}~\bibnamefont {Del~Sole}}, \
  and\ \bibinfo {author} {\bibfnamefont {A.}~\bibnamefont {Marini}},\
  }\href@noop {} {\bibfield  {journal} {\bibinfo  {journal} {Phys. Rev. B}\
  }\textbf {\bibinfo {volume} {68}},\ \bibinfo {pages} {165108} (\bibinfo
  {year} {2003})}\BibitemShut {NoStop}%
\bibitem [{\citenamefont {Marini}\ \emph {et~al.}(2003)\citenamefont {Marini},
  \citenamefont {Del~Sole},\ and\ \citenamefont {Rubio}}]{MDR03}%
  \BibitemOpen
  \bibfield  {author} {\bibinfo {author} {\bibfnamefont {A.}~\bibnamefont
  {Marini}}, \bibinfo {author} {\bibfnamefont {R.}~\bibnamefont {Del~Sole}}, \
  and\ \bibinfo {author} {\bibfnamefont {A.}~\bibnamefont {Rubio}},\
  }\href@noop {} {\bibfield  {journal} {\bibinfo  {journal} {Phys. Rev. Lett.}\
  }\textbf {\bibinfo {volume} {91}},\ \bibinfo {pages} {256402} (\bibinfo
  {year} {2003})}\BibitemShut {NoStop}%
\bibitem [{\citenamefont {Stephens}\ \emph {et~al.}(1994)\citenamefont
  {Stephens}, \citenamefont {Devlin}, \citenamefont {Chabalowski},\ and\
  \citenamefont {Frisch}}]{SDCF94}%
  \BibitemOpen
  \bibfield  {author} {\bibinfo {author} {\bibfnamefont {P.~J.}\ \bibnamefont
  {Stephens}}, \bibinfo {author} {\bibfnamefont {F.~J.}\ \bibnamefont
  {Devlin}}, \bibinfo {author} {\bibfnamefont {C.~F.}\ \bibnamefont
  {Chabalowski}}, \ and\ \bibinfo {author} {\bibfnamefont {M.~J.}\ \bibnamefont
  {Frisch}},\ }\href@noop {} {\bibfield  {journal} {\bibinfo  {journal} {J.
  Phys. Chem.}\ }\textbf {\bibinfo {volume} {98}},\ \bibinfo {pages} {11623}
  (\bibinfo {year} {1994})}\BibitemShut {NoStop}%
\bibitem [{\citenamefont {Bernasconi}\ \emph {et~al.}(2011)\citenamefont
  {Bernasconi}, \citenamefont {Tomi\'c}, \citenamefont {Ferrero}, \citenamefont
  {R\'erat}, \citenamefont {Orlando}, \citenamefont {Dovesi},\ and\
  \citenamefont {Harrison}}]{Bernasconi2011}%
  \BibitemOpen
  \bibfield  {author} {\bibinfo {author} {\bibfnamefont {L.}~\bibnamefont
  {Bernasconi}}, \bibinfo {author} {\bibfnamefont {S.}~\bibnamefont {Tomi\'c}},
  \bibinfo {author} {\bibfnamefont {M.}~\bibnamefont {Ferrero}}, \bibinfo
  {author} {\bibfnamefont {M.}~\bibnamefont {R\'erat}}, \bibinfo {author}
  {\bibfnamefont {R.}~\bibnamefont {Orlando}}, \bibinfo {author} {\bibfnamefont
  {R.}~\bibnamefont {Dovesi}}, \ and\ \bibinfo {author} {\bibfnamefont {N.~M.}\
  \bibnamefont {Harrison}},\ }\href@noop {} {\bibfield  {journal} {\bibinfo
  {journal} {Phys. Rev. B}\ }\textbf {\bibinfo {volume} {83}},\ \bibinfo
  {pages} {195325} (\bibinfo {year} {2011})}\BibitemShut {NoStop}%
\bibitem [{\citenamefont {Tomi\'c}\ \emph {et~al.}(2014)\citenamefont
  {Tomi\'c}, \citenamefont {Bernasconi}, \citenamefont {Searle},\ and\
  \citenamefont {Harrison}}]{Tomic2014}%
  \BibitemOpen
  \bibfield  {author} {\bibinfo {author} {\bibfnamefont {S.}~\bibnamefont
  {Tomi\'c}}, \bibinfo {author} {\bibfnamefont {L.}~\bibnamefont {Bernasconi}},
  \bibinfo {author} {\bibfnamefont {B.~G.}\ \bibnamefont {Searle}}, \ and\
  \bibinfo {author} {\bibfnamefont {N.~M.}\ \bibnamefont {Harrison}},\
  }\href@noop {} {\bibfield  {journal} {\bibinfo  {journal} {J. Phys. Chem. C}\
  }\textbf {\bibinfo {volume} {118}},\ \bibinfo {pages} {14478} (\bibinfo
  {year} {2014})}\BibitemShut {NoStop}%
\bibitem [{\citenamefont {Refaely-Abramson}\ \emph {et~al.}(2015)\citenamefont
  {Refaely-Abramson}, \citenamefont {Jain}, \citenamefont {Sharifzadeh},
  \citenamefont {Neaton},\ and\ \citenamefont {Kronik}}]{Refaely2015}%
  \BibitemOpen
  \bibfield  {author} {\bibinfo {author} {\bibfnamefont {S.}~\bibnamefont
  {Refaely-Abramson}}, \bibinfo {author} {\bibfnamefont {M.}~\bibnamefont
  {Jain}}, \bibinfo {author} {\bibfnamefont {S.}~\bibnamefont {Sharifzadeh}},
  \bibinfo {author} {\bibfnamefont {J.~B.}\ \bibnamefont {Neaton}}, \ and\
  \bibinfo {author} {\bibfnamefont {L.}~\bibnamefont {Kronik}},\ }\href@noop {}
  {\bibfield  {journal} {\bibinfo  {journal} {Phys. Rev. B}\ }\textbf {\bibinfo
  {volume} {92}},\ \bibinfo {pages} {081204} (\bibinfo {year}
  {2015})}\BibitemShut {NoStop}%
\bibitem [{\citenamefont {Yang}\ \emph {et~al.}(2015)\citenamefont {Yang},
  \citenamefont {Sottile},\ and\ \citenamefont {Ullrich}}]{Yang2015}%
  \BibitemOpen
  \bibfield  {author} {\bibinfo {author} {\bibfnamefont {Z.-H.}\ \bibnamefont
  {Yang}}, \bibinfo {author} {\bibfnamefont {F.}~\bibnamefont {Sottile}}, \
  and\ \bibinfo {author} {\bibfnamefont {C.~A.}\ \bibnamefont {Ullrich}},\
  }\href@noop {} {\bibfield  {journal} {\bibinfo  {journal} {Phys. Rev. B}\
  }\textbf {\bibinfo {volume} {92}},\ \bibinfo {pages} {035202} (\bibinfo
  {year} {2015})}\BibitemShut {NoStop}%
\bibitem [{\citenamefont {Botti}\ \emph {et~al.}(2004)\citenamefont {Botti},
  \citenamefont {Sottile}, \citenamefont {Vast}, \citenamefont {Olevano},
  \citenamefont {Reining}, \citenamefont {Weissker}, \citenamefont {Rubio},
  \citenamefont {Onida}, \citenamefont {Del~Sole},\ and\ \citenamefont
  {Godby}}]{Botti04}%
  \BibitemOpen
  \bibfield  {author} {\bibinfo {author} {\bibfnamefont {S.}~\bibnamefont
  {Botti}}, \bibinfo {author} {\bibfnamefont {F.}~\bibnamefont {Sottile}},
  \bibinfo {author} {\bibfnamefont {N.}~\bibnamefont {Vast}}, \bibinfo {author}
  {\bibfnamefont {V.}~\bibnamefont {Olevano}}, \bibinfo {author} {\bibfnamefont
  {L.}~\bibnamefont {Reining}}, \bibinfo {author} {\bibfnamefont {H.-C.}\
  \bibnamefont {Weissker}}, \bibinfo {author} {\bibfnamefont {A.}~\bibnamefont
  {Rubio}}, \bibinfo {author} {\bibfnamefont {G.}~\bibnamefont {Onida}},
  \bibinfo {author} {\bibfnamefont {R.}~\bibnamefont {Del~Sole}}, \ and\
  \bibinfo {author} {\bibfnamefont {R.~W.}\ \bibnamefont {Godby}},\ }\href@noop
  {} {\bibfield  {journal} {\bibinfo  {journal} {Phys. Rev. B}\ }\textbf
  {\bibinfo {volume} {69}},\ \bibinfo {pages} {155112} (\bibinfo {year}
  {2004})}\BibitemShut {NoStop}%
\bibitem [{\citenamefont {Botti}\ \emph {et~al.}(2007)\citenamefont {Botti},
  \citenamefont {Schindlmayr}, \citenamefont {{Del Sole}},\ and\ \citenamefont
  {Reining}}]{Botti2007}%
  \BibitemOpen
  \bibfield  {author} {\bibinfo {author} {\bibfnamefont {S.}~\bibnamefont
  {Botti}}, \bibinfo {author} {\bibfnamefont {A.}~\bibnamefont {Schindlmayr}},
  \bibinfo {author} {\bibfnamefont {R.}~\bibnamefont {{Del Sole}}}, \ and\
  \bibinfo {author} {\bibfnamefont {L.}~\bibnamefont {Reining}},\ }\href@noop
  {} {\bibfield  {journal} {\bibinfo  {journal} {Rep. Prog. Phys.}\ }\textbf
  {\bibinfo {volume} {70}},\ \bibinfo {pages} {357} (\bibinfo {year}
  {2007})}\BibitemShut {NoStop}%
\bibitem [{\citenamefont {Sharma}\ \emph {et~al.}(2011)\citenamefont {Sharma},
  \citenamefont {Dewhurst}, \citenamefont {Sanna},\ and\ \citenamefont
  {Gross}}]{Sharma11}%
  \BibitemOpen
  \bibfield  {author} {\bibinfo {author} {\bibfnamefont {S.}~\bibnamefont
  {Sharma}}, \bibinfo {author} {\bibfnamefont {J.~K.}\ \bibnamefont
  {Dewhurst}}, \bibinfo {author} {\bibfnamefont {A.}~\bibnamefont {Sanna}}, \
  and\ \bibinfo {author} {\bibfnamefont {E.~K.~U.}\ \bibnamefont {Gross}},\
  }\href@noop {} {\bibfield  {journal} {\bibinfo  {journal} {Phys. Rev. Lett.}\
  }\textbf {\bibinfo {volume} {107}},\ \bibinfo {pages} {186401} (\bibinfo
  {year} {2011})}\BibitemShut {NoStop}%
\bibitem [{\citenamefont {Rigamonti}\ \emph {et~al.}(2015)\citenamefont
  {Rigamonti}, \citenamefont {Botti}, \citenamefont {Veniard}, \citenamefont
  {Draxl}, \citenamefont {Reining},\ and\ \citenamefont
  {Sottile}}]{Rigamonti15}%
  \BibitemOpen
  \bibfield  {author} {\bibinfo {author} {\bibfnamefont {S.}~\bibnamefont
  {Rigamonti}}, \bibinfo {author} {\bibfnamefont {S.}~\bibnamefont {Botti}},
  \bibinfo {author} {\bibfnamefont {V.}~\bibnamefont {Veniard}}, \bibinfo
  {author} {\bibfnamefont {C.}~\bibnamefont {Draxl}}, \bibinfo {author}
  {\bibfnamefont {L.}~\bibnamefont {Reining}}, \ and\ \bibinfo {author}
  {\bibfnamefont {F.}~\bibnamefont {Sottile}},\ }\href@noop {} {\bibfield
  {journal} {\bibinfo  {journal} {Phys. Rev. Lett.}\ }\textbf {\bibinfo
  {volume} {114}},\ \bibinfo {pages} {146402} (\bibinfo {year}
  {2015})}\BibitemShut {NoStop}%
\bibitem [{\citenamefont {Trevisanutto}\ \emph {et~al.}(2013)\citenamefont
  {Trevisanutto}, \citenamefont {Terentjevs}, \citenamefont {Constantin},
  \citenamefont {Olevano},\ and\ \citenamefont {Sala}}]{Trevisanutto13}%
  \BibitemOpen
  \bibfield  {author} {\bibinfo {author} {\bibfnamefont {P.~E.}\ \bibnamefont
  {Trevisanutto}}, \bibinfo {author} {\bibfnamefont {A.}~\bibnamefont
  {Terentjevs}}, \bibinfo {author} {\bibfnamefont {L.~A.}\ \bibnamefont
  {Constantin}}, \bibinfo {author} {\bibfnamefont {V.}~\bibnamefont {Olevano}},
  \ and\ \bibinfo {author} {\bibfnamefont {F.~D.}\ \bibnamefont {Sala}},\
  }\href@noop {} {\bibfield  {journal} {\bibinfo  {journal} {Phys. Rev. B}\
  }\textbf {\bibinfo {volume} {87}},\ \bibinfo {pages} {205143} (\bibinfo
  {year} {2013})}\BibitemShut {NoStop}%
\bibitem [{\citenamefont {Sharma}\ \emph {et~al.}(2016)\citenamefont {Sharma},
  \citenamefont {Dewhurst}, \citenamefont {Sanna},\ and\ \citenamefont
  {Gross}}]{Sharma16}%
  \BibitemOpen
  \bibfield  {author} {\bibinfo {author} {\bibfnamefont {S.}~\bibnamefont
  {Sharma}}, \bibinfo {author} {\bibfnamefont {J.~K.}\ \bibnamefont
  {Dewhurst}}, \bibinfo {author} {\bibfnamefont {A.}~\bibnamefont {Sanna}}, \
  and\ \bibinfo {author} {\bibfnamefont {E.~K.~U.}\ \bibnamefont {Gross}},\
  }\href@noop {} {\bibfield  {journal} {\bibinfo  {journal} {Phys. Rev. Lett.}\
  }\textbf {\bibinfo {volume} {117}},\ \bibinfo {pages} {159701} (\bibinfo
  {year} {2016})}\BibitemShut {NoStop}%
\bibitem [{\citenamefont {Rigamonti}\ \emph {et~al.}(2016)\citenamefont
  {Rigamonti}, \citenamefont {Botti}, \citenamefont {Veniard}, \citenamefont
  {Draxl}, \citenamefont {Reining},\ and\ \citenamefont
  {Sottile}}]{Rigamonti16}%
  \BibitemOpen
  \bibfield  {author} {\bibinfo {author} {\bibfnamefont {S.}~\bibnamefont
  {Rigamonti}}, \bibinfo {author} {\bibfnamefont {S.}~\bibnamefont {Botti}},
  \bibinfo {author} {\bibfnamefont {V.}~\bibnamefont {Veniard}}, \bibinfo
  {author} {\bibfnamefont {C.}~\bibnamefont {Draxl}}, \bibinfo {author}
  {\bibfnamefont {L.}~\bibnamefont {Reining}}, \ and\ \bibinfo {author}
  {\bibfnamefont {F.}~\bibnamefont {Sottile}},\ }\href@noop {} {\bibfield
  {journal} {\bibinfo  {journal} {Phys. Rev. Lett.}\ }\textbf {\bibinfo
  {volume} {117}},\ \bibinfo {pages} {159702} (\bibinfo {year}
  {2016})}\BibitemShut {NoStop}%
\bibitem [{\citenamefont {Botti}\ \emph {et~al.}(2005)\citenamefont {Botti},
  \citenamefont {Fourreau}, \citenamefont {Nguyen}, \citenamefont {Renault},
  \citenamefont {Sottile},\ and\ \citenamefont {Reining}}]{Botti05}%
  \BibitemOpen
  \bibfield  {author} {\bibinfo {author} {\bibfnamefont {S.}~\bibnamefont
  {Botti}}, \bibinfo {author} {\bibfnamefont {A.}~\bibnamefont {Fourreau}},
  \bibinfo {author} {\bibfnamefont {F.}~\bibnamefont {Nguyen}}, \bibinfo
  {author} {\bibfnamefont {Y.-O.}\ \bibnamefont {Renault}}, \bibinfo {author}
  {\bibfnamefont {F.}~\bibnamefont {Sottile}}, \ and\ \bibinfo {author}
  {\bibfnamefont {L.}~\bibnamefont {Reining}},\ }\href@noop {} {\bibfield
  {journal} {\bibinfo  {journal} {Phys. Rev. B}\ }\textbf {\bibinfo {volume}
  {72}},\ \bibinfo {pages} {125203} (\bibinfo {year} {2005})}\BibitemShut
  {NoStop}%
\bibitem [{\citenamefont {Berger}(2015)}]{Berger15}%
  \BibitemOpen
  \bibfield  {author} {\bibinfo {author} {\bibfnamefont {J.~A.}\ \bibnamefont
  {Berger}},\ }\href@noop {} {\bibfield  {journal} {\bibinfo  {journal} {Phys.
  Rev. Lett.}\ }\textbf {\bibinfo {volume} {115}},\ \bibinfo {pages} {137402}
  (\bibinfo {year} {2015})}\BibitemShut {NoStop}%
\bibitem [{\citenamefont {Gajdo\ifmmode~\check{s}\else \v{s}\fi{}}\ \emph
  {et~al.}(2006)\citenamefont {Gajdo\ifmmode~\check{s}\else \v{s}\fi{}},
  \citenamefont {Hummer}, \citenamefont {Kresse}, \citenamefont
  {Furthm\"uller},\ and\ \citenamefont {Bechstedt}}]{Gajdos06}%
  \BibitemOpen
  \bibfield  {author} {\bibinfo {author} {\bibfnamefont {M.}~\bibnamefont
  {Gajdo\ifmmode~\check{s}\else \v{s}\fi{}}}, \bibinfo {author} {\bibfnamefont
  {K.}~\bibnamefont {Hummer}}, \bibinfo {author} {\bibfnamefont
  {G.}~\bibnamefont {Kresse}}, \bibinfo {author} {\bibfnamefont
  {J.}~\bibnamefont {Furthm\"uller}}, \ and\ \bibinfo {author} {\bibfnamefont
  {F.}~\bibnamefont {Bechstedt}},\ }\href@noop {} {\bibfield  {journal}
  {\bibinfo  {journal} {Phys. Rev. B}\ }\textbf {\bibinfo {volume} {73}},\
  \bibinfo {pages} {045112} (\bibinfo {year} {2006})}\BibitemShut {NoStop}%
\bibitem [{\citenamefont {Baroni}\ and\ \citenamefont
  {Resta}(1986)}]{Baroni86}%
  \BibitemOpen
  \bibfield  {author} {\bibinfo {author} {\bibfnamefont {S.}~\bibnamefont
  {Baroni}}\ and\ \bibinfo {author} {\bibfnamefont {R.}~\bibnamefont {Resta}},\
  }\href@noop {} {\bibfield  {journal} {\bibinfo  {journal} {Phys. Rev. B}\
  }\textbf {\bibinfo {volume} {33}},\ \bibinfo {pages} {7017} (\bibinfo {year}
  {1986})}\BibitemShut {NoStop}%
\bibitem [{\citenamefont {Casida}(1995)}]{Casida1995}%
  \BibitemOpen
  \bibfield  {author} {\bibinfo {author} {\bibfnamefont {M.~E.}\ \bibnamefont
  {Casida}},\ }in\ \href@noop {} {\emph {\bibinfo {booktitle} {Recent Advances
  in Density Functional Methods}}},\ \bibinfo {series} {Recent Advances in
  Computational Chemistry}, Vol.~\bibinfo {volume} {1},\ \bibinfo {editor}
  {edited by\ \bibinfo {editor} {\bibfnamefont {D.~E.}\ \bibnamefont {Chong}}}\
  (\bibinfo  {publisher} {World Scientific},\ \bibinfo {address} {Singapore},\
  \bibinfo {year} {1995})\ pp.\ \bibinfo {pages} {155--92}\BibitemShut
  {NoStop}%
\bibitem [{\citenamefont {Yang}\ \emph {et~al.}(2012)\citenamefont {Yang},
  \citenamefont {Li},\ and\ \citenamefont {Ullrich}}]{Yang2012}%
  \BibitemOpen
  \bibfield  {author} {\bibinfo {author} {\bibfnamefont {Z.-H.}\ \bibnamefont
  {Yang}}, \bibinfo {author} {\bibfnamefont {Y.}~\bibnamefont {Li}}, \ and\
  \bibinfo {author} {\bibfnamefont {C.~A.}\ \bibnamefont {Ullrich}},\
  }\href@noop {} {\bibfield  {journal} {\bibinfo  {journal} {J. Chem. Phys.}\
  }\textbf {\bibinfo {volume} {137}},\ \bibinfo {pages} {014513} (\bibinfo
  {year} {2012})}\BibitemShut {NoStop}%
\bibitem [{\citenamefont {Yang}\ and\ \citenamefont
  {Ullrich}(2013)}]{Yang2013}%
  \BibitemOpen
  \bibfield  {author} {\bibinfo {author} {\bibfnamefont {Z.-H.}\ \bibnamefont
  {Yang}}\ and\ \bibinfo {author} {\bibfnamefont {C.~A.}\ \bibnamefont
  {Ullrich}},\ }\href@noop {} {\bibfield  {journal} {\bibinfo  {journal} {Phys.
  Rev. B}\ }\textbf {\bibinfo {volume} {87}},\ \bibinfo {pages} {195204}
  (\bibinfo {year} {2013})}\BibitemShut {NoStop}%
\bibitem [{\citenamefont {Byun}\ and\ \citenamefont
  {Ullrich}(2017)}]{Byun2017}%
  \BibitemOpen
  \bibfield  {author} {\bibinfo {author} {\bibfnamefont {Y.-M.}\ \bibnamefont
  {Byun}}\ and\ \bibinfo {author} {\bibfnamefont {C.~A.}\ \bibnamefont
  {Ullrich}},\ }\href@noop {} {\bibfield  {journal} {\bibinfo  {journal}
  {Computation}\ }\textbf {\bibinfo {volume} {5}},\ \bibinfo {pages} {9}
  (\bibinfo {year} {2017})}\BibitemShut {NoStop}%
\bibitem [{\citenamefont {{See Supplemental Material at
  http://...}}()}]{supplemental}%
  \BibitemOpen
  \bibfield  {author} {\bibinfo {author} {\bibnamefont {{See Supplemental
  Material at http://...}}},\ }\href@noop {} {\ }\BibitemShut {NoStop}%
\bibitem [{\citenamefont {Sharma}\ \emph {et~al.}(2012)\citenamefont {Sharma},
  \citenamefont {Dewhurst}, \citenamefont {Sanna}, \citenamefont {Rubio},\ and\
  \citenamefont {Gross}}]{Sharma2012}%
  \BibitemOpen
  \bibfield  {author} {\bibinfo {author} {\bibfnamefont {S.}~\bibnamefont
  {Sharma}}, \bibinfo {author} {\bibfnamefont {J.~K.}\ \bibnamefont
  {Dewhurst}}, \bibinfo {author} {\bibfnamefont {A.}~\bibnamefont {Sanna}},
  \bibinfo {author} {\bibfnamefont {A.}~\bibnamefont {Rubio}}, \ and\ \bibinfo
  {author} {\bibfnamefont {E.~K.~U.}\ \bibnamefont {Gross}},\ }\href@noop {}
  {\bibfield  {journal} {\bibinfo  {journal} {New J. Phys.}\ }\textbf {\bibinfo
  {volume} {14}},\ \bibinfo {pages} {053052} (\bibinfo {year}
  {2012})}\BibitemShut {NoStop}%
\bibitem [{\citenamefont {Sharma}()}]{Sharma_commun}%
  \BibitemOpen
  \bibfield  {author} {\bibinfo {author} {\bibfnamefont {S.}~\bibnamefont
  {Sharma}},\ }\href@noop {} {\bibinfo  {journal} {private communication}\
  }\BibitemShut {NoStop}%
\bibitem [{\citenamefont {Constantin}\ and\ \citenamefont
  {Pitarke}(2007)}]{Constantin07}%
  \BibitemOpen
\bibfield  {journal} {  }\bibfield  {author} {\bibinfo {author} {\bibfnamefont
  {L.~A.}\ \bibnamefont {Constantin}}\ and\ \bibinfo {author} {\bibfnamefont
  {J.~M.}\ \bibnamefont {Pitarke}},\ }\href@noop {} {\bibfield  {journal}
  {\bibinfo  {journal} {Phys. Rev. B}\ }\textbf {\bibinfo {volume} {75}},\
  \bibinfo {pages} {245127} (\bibinfo {year} {2007})}\BibitemShut {NoStop}%
\bibitem [{\citenamefont {Gonze}\ \emph {et~al.}(2009)\citenamefont {Gonze},
  \citenamefont {Amadon}, \citenamefont {Anglade}, \citenamefont {Beuken},
  \citenamefont {Bottin}, \citenamefont {Boulanger}, \citenamefont {Bruneval},
  \citenamefont {Caliste}, \citenamefont {Caracas}, \citenamefont
  {C{\^o}t{\'e}}, \citenamefont {Deutsch}, \citenamefont {Genovese},
  \citenamefont {Ghosez}, \citenamefont {Giantomassi}, \citenamefont
  {Goedecker}, \citenamefont {Hamann}, \citenamefont {Hermet}, \citenamefont
  {Jollet}, \citenamefont {Jomard}, \citenamefont {Leroux}, \citenamefont
  {Mancini}, \citenamefont {Mazevet}, \citenamefont {Oliveira}, \citenamefont
  {Onida}, \citenamefont {Pouillon}, \citenamefont {Rangel}, \citenamefont
  {Rignanese}, \citenamefont {Sangalli}, \citenamefont {Shaltaf}, \citenamefont
  {Torrent}, \citenamefont {Verstraete}, \citenamefont {Zerah},\ and\
  \citenamefont {Zwanziger}}]{Gonze09}%
  \BibitemOpen
  \bibfield  {author} {\bibinfo {author} {\bibfnamefont {X.}~\bibnamefont
  {Gonze}}, \bibinfo {author} {\bibfnamefont {B.}~\bibnamefont {Amadon}},
  \bibinfo {author} {\bibfnamefont {P.-M.}\ \bibnamefont {Anglade}}, \bibinfo
  {author} {\bibfnamefont {J.-M.}\ \bibnamefont {Beuken}}, \bibinfo {author}
  {\bibfnamefont {F.}~\bibnamefont {Bottin}}, \bibinfo {author} {\bibfnamefont
  {P.}~\bibnamefont {Boulanger}}, \bibinfo {author} {\bibfnamefont
  {F.}~\bibnamefont {Bruneval}}, \bibinfo {author} {\bibfnamefont
  {D.}~\bibnamefont {Caliste}}, \bibinfo {author} {\bibfnamefont
  {R.}~\bibnamefont {Caracas}}, \bibinfo {author} {\bibfnamefont
  {M.}~\bibnamefont {C{\^o}t{\'e}}}, \bibinfo {author} {\bibfnamefont
  {T.}~\bibnamefont {Deutsch}}, \bibinfo {author} {\bibfnamefont
  {L.}~\bibnamefont {Genovese}}, \bibinfo {author} {\bibfnamefont
  {P.}~\bibnamefont {Ghosez}}, \bibinfo {author} {\bibfnamefont
  {M.}~\bibnamefont {Giantomassi}}, \bibinfo {author} {\bibfnamefont
  {S.}~\bibnamefont {Goedecker}}, \bibinfo {author} {\bibfnamefont
  {D.}~\bibnamefont {Hamann}}, \bibinfo {author} {\bibfnamefont
  {P.}~\bibnamefont {Hermet}}, \bibinfo {author} {\bibfnamefont
  {F.}~\bibnamefont {Jollet}}, \bibinfo {author} {\bibfnamefont
  {G.}~\bibnamefont {Jomard}}, \bibinfo {author} {\bibfnamefont
  {S.}~\bibnamefont {Leroux}}, \bibinfo {author} {\bibfnamefont
  {M.}~\bibnamefont {Mancini}}, \bibinfo {author} {\bibfnamefont
  {S.}~\bibnamefont {Mazevet}}, \bibinfo {author} {\bibfnamefont
  {M.}~\bibnamefont {Oliveira}}, \bibinfo {author} {\bibfnamefont
  {G.}~\bibnamefont {Onida}}, \bibinfo {author} {\bibfnamefont
  {Y.}~\bibnamefont {Pouillon}}, \bibinfo {author} {\bibfnamefont
  {T.}~\bibnamefont {Rangel}}, \bibinfo {author} {\bibfnamefont {G.-M.}\
  \bibnamefont {Rignanese}}, \bibinfo {author} {\bibfnamefont {D.}~\bibnamefont
  {Sangalli}}, \bibinfo {author} {\bibfnamefont {R.}~\bibnamefont {Shaltaf}},
  \bibinfo {author} {\bibfnamefont {M.}~\bibnamefont {Torrent}}, \bibinfo
  {author} {\bibfnamefont {M.}~\bibnamefont {Verstraete}}, \bibinfo {author}
  {\bibfnamefont {G.}~\bibnamefont {Zerah}}, \ and\ \bibinfo {author}
  {\bibfnamefont {J.}~\bibnamefont {Zwanziger}},\ }\href@noop {} {\bibfield
  {journal} {\bibinfo  {journal} {Comput. Phys. Commun.}\ }\textbf {\bibinfo
  {volume} {180}},\ \bibinfo {pages} {2582 } (\bibinfo {year}
  {2009})}\BibitemShut {NoStop}%
\bibitem [{\citenamefont {Olevano}\ \emph {et~al.}()\citenamefont {Olevano},
  \citenamefont {Reining},\ and\ \citenamefont {Sottile}}]{Olevano97}%
  \BibitemOpen
  \bibfield  {author} {\bibinfo {author} {\bibfnamefont {V.}~\bibnamefont
  {Olevano}}, \bibinfo {author} {\bibfnamefont {L.}~\bibnamefont {Reining}}, \
  and\ \bibinfo {author} {\bibfnamefont {F.}~\bibnamefont {Sottile}},\
  }\href@noop {} {\bibinfo  {journal} {http://www.dp-code.org/}\ }\BibitemShut
  {NoStop}%
\bibitem [{\citenamefont {Cazzaniga}\ \emph {et~al.}(2011)\citenamefont
  {Cazzaniga}, \citenamefont {Weissker}, \citenamefont {Huotari}, \citenamefont
  {Pylkk\"anen}, \citenamefont {Salvestrini}, \citenamefont {Monaco},
  \citenamefont {Onida},\ and\ \citenamefont {Reining}}]{Cazzaniga11}%
  \BibitemOpen
\bibfield  {journal} {  }\bibfield  {author} {\bibinfo {author} {\bibfnamefont
  {M.}~\bibnamefont {Cazzaniga}}, \bibinfo {author} {\bibfnamefont {H.-C.}\
  \bibnamefont {Weissker}}, \bibinfo {author} {\bibfnamefont {S.}~\bibnamefont
  {Huotari}}, \bibinfo {author} {\bibfnamefont {T.}~\bibnamefont
  {Pylkk\"anen}}, \bibinfo {author} {\bibfnamefont {P.}~\bibnamefont
  {Salvestrini}}, \bibinfo {author} {\bibfnamefont {G.}~\bibnamefont {Monaco}},
  \bibinfo {author} {\bibfnamefont {G.}~\bibnamefont {Onida}}, \ and\ \bibinfo
  {author} {\bibfnamefont {L.}~\bibnamefont {Reining}},\ }\href@noop {}
  {\bibfield  {journal} {\bibinfo  {journal} {Phys. Rev. B}\ }\textbf {\bibinfo
  {volume} {84}},\ \bibinfo {pages} {075109} (\bibinfo {year}
  {2011})}\BibitemShut {NoStop}%
\bibitem [{\citenamefont {Stubner}\ \emph {et~al.}(2004)\citenamefont
  {Stubner}, \citenamefont {Tokatly},\ and\ \citenamefont
  {Pankratov}}]{Stubner2004}%
  \BibitemOpen
  \bibfield  {author} {\bibinfo {author} {\bibfnamefont {R.}~\bibnamefont
  {Stubner}}, \bibinfo {author} {\bibfnamefont {I.~V.}\ \bibnamefont
  {Tokatly}}, \ and\ \bibinfo {author} {\bibfnamefont {O.}~\bibnamefont
  {Pankratov}},\ }\href@noop {} {\bibfield  {journal} {\bibinfo  {journal}
  {Phys. Rev. B}\ }\textbf {\bibinfo {volume} {70}},\ \bibinfo {pages} {245119}
  (\bibinfo {year} {2004})}\BibitemShut {NoStop}%
\bibitem [{\citenamefont {Sharma}\ \emph {et~al.}(2014)\citenamefont {Sharma},
  \citenamefont {Dewhurst},\ and\ \citenamefont {Gross}}]{Sharma2014a}%
  \BibitemOpen
  \bibfield  {author} {\bibinfo {author} {\bibfnamefont {S.}~\bibnamefont
  {Sharma}}, \bibinfo {author} {\bibfnamefont {J.~K.}\ \bibnamefont
  {Dewhurst}}, \ and\ \bibinfo {author} {\bibfnamefont {E.~K.~U.}\ \bibnamefont
  {Gross}},\ }in\ \href@noop {} {\emph {\bibinfo {booktitle} {First Principles
  Approaches to Spectroscopic Properties of Complex Materials}}},\ \bibinfo
  {series} {Topics in Current Chemistry}, Vol.\ \bibinfo {volume} {347},\
  \bibinfo {editor} {edited by\ \bibinfo {editor} {\bibfnamefont {C.~D.}\
  \bibnamefont {Valentin}}, \bibinfo {editor} {\bibfnamefont {S.}~\bibnamefont
  {Botti}}, \ and\ \bibinfo {editor} {\bibfnamefont {M.}~\bibnamefont
  {Cococcioni}}}\ (\bibinfo  {publisher} {Springer},\ \bibinfo {address}
  {Berlin},\ \bibinfo {year} {2014})\ p.\ \bibinfo {pages} {235}\BibitemShut
  {NoStop}%
\bibitem [{\citenamefont {Levine}\ and\ \citenamefont
  {Allan}(1989)}]{Levine89}%
  \BibitemOpen
  \bibfield  {author} {\bibinfo {author} {\bibfnamefont {Z.~H.}\ \bibnamefont
  {Levine}}\ and\ \bibinfo {author} {\bibfnamefont {D.~C.}\ \bibnamefont
  {Allan}},\ }\href@noop {} {\bibfield  {journal} {\bibinfo  {journal} {Phys.
  Rev. Lett.}\ }\textbf {\bibinfo {volume} {63}},\ \bibinfo {pages} {1719}
  (\bibinfo {year} {1989})}\BibitemShut {NoStop}%
\bibitem [{\citenamefont {Gonze}\ and\ \citenamefont {Lee}(1997)}]{Gonze97}%
  \BibitemOpen
  \bibfield  {author} {\bibinfo {author} {\bibfnamefont {X.}~\bibnamefont
  {Gonze}}\ and\ \bibinfo {author} {\bibfnamefont {C.}~\bibnamefont {Lee}},\
  }\href@noop {} {\bibfield  {journal} {\bibinfo  {journal} {Phys. Rev. B}\
  }\textbf {\bibinfo {volume} {55}},\ \bibinfo {pages} {10355} (\bibinfo {year}
  {1997})}\BibitemShut {NoStop}%
\bibitem [{\citenamefont {Del~Sole}\ and\ \citenamefont
  {Girlanda}(1993)}]{DelSole93}%
  \BibitemOpen
  \bibfield  {author} {\bibinfo {author} {\bibfnamefont {R.}~\bibnamefont
  {Del~Sole}}\ and\ \bibinfo {author} {\bibfnamefont {R.}~\bibnamefont
  {Girlanda}},\ }\href@noop {} {\bibfield  {journal} {\bibinfo  {journal}
  {Phys. Rev. B}\ }\textbf {\bibinfo {volume} {48}},\ \bibinfo {pages} {11789}
  (\bibinfo {year} {1993})}\BibitemShut {NoStop}%
\bibitem [{\citenamefont {Baroni}\ \emph {et~al.}(2001)\citenamefont {Baroni},
  \citenamefont {de~Gironcoli}, \citenamefont {Dal~Corso},\ and\ \citenamefont
  {Giannozzi}}]{Baroni01}%
  \BibitemOpen
  \bibfield  {author} {\bibinfo {author} {\bibfnamefont {S.}~\bibnamefont
  {Baroni}}, \bibinfo {author} {\bibfnamefont {S.}~\bibnamefont
  {de~Gironcoli}}, \bibinfo {author} {\bibfnamefont {A.}~\bibnamefont
  {Dal~Corso}}, \ and\ \bibinfo {author} {\bibfnamefont {P.}~\bibnamefont
  {Giannozzi}},\ }\href@noop {} {\bibfield  {journal} {\bibinfo  {journal}
  {Rev. Mod. Phys.}\ }\textbf {\bibinfo {volume} {73}},\ \bibinfo {pages} {515}
  (\bibinfo {year} {2001})}\BibitemShut {NoStop}%
\bibitem [{\citenamefont {Parenteau}\ \emph {et~al.}(1992)\citenamefont
  {Parenteau}, \citenamefont {Carlone},\ and\ \citenamefont {Khanna}}]{PCK92}%
  \BibitemOpen
  \bibfield  {author} {\bibinfo {author} {\bibfnamefont {M.}~\bibnamefont
  {Parenteau}}, \bibinfo {author} {\bibfnamefont {C.}~\bibnamefont {Carlone}},
  \ and\ \bibinfo {author} {\bibfnamefont {S.~M.}\ \bibnamefont {Khanna}},\
  }\href@noop {} {\bibfield  {journal} {\bibinfo  {journal} {J. Appl. Phys.}\
  }\textbf {\bibinfo {volume} {71}},\ \bibinfo {pages} {3747} (\bibinfo {year}
  {1992})}\BibitemShut {NoStop}%
\bibitem [{\citenamefont {As}\ \emph {et~al.}(1997)\citenamefont {As},
  \citenamefont {Schmilgus}, \citenamefont {Wang}, \citenamefont
  {Sch\"{o}ttker}, \citenamefont {Schikora},\ and\ \citenamefont
  {Lischka}}]{ASWS97}%
  \BibitemOpen
  \bibfield  {author} {\bibinfo {author} {\bibfnamefont {D.~J.}\ \bibnamefont
  {As}}, \bibinfo {author} {\bibfnamefont {F.}~\bibnamefont {Schmilgus}},
  \bibinfo {author} {\bibfnamefont {C.}~\bibnamefont {Wang}}, \bibinfo {author}
  {\bibfnamefont {B.}~\bibnamefont {Sch\"{o}ttker}}, \bibinfo {author}
  {\bibfnamefont {D.}~\bibnamefont {Schikora}}, \ and\ \bibinfo {author}
  {\bibfnamefont {K.}~\bibnamefont {Lischka}},\ }\href@noop {} {\bibfield
  {journal} {\bibinfo  {journal} {Appl. Phys. Lett.}\ }\textbf {\bibinfo
  {volume} {70}},\ \bibinfo {pages} {1311} (\bibinfo {year}
  {1997})}\BibitemShut {NoStop}%
\bibitem [{\citenamefont {Muth}\ \emph {et~al.}(1997)\citenamefont {Muth},
  \citenamefont {Lee}, \citenamefont {Shmagin}, \citenamefont {Kolbas},
  \citenamefont {Casey}, \citenamefont {Keller}, \citenamefont {Mishra},\ and\
  \citenamefont {DenBaars}}]{MLSK97}%
  \BibitemOpen
  \bibfield  {author} {\bibinfo {author} {\bibfnamefont {J.~F.}\ \bibnamefont
  {Muth}}, \bibinfo {author} {\bibfnamefont {J.~H.}\ \bibnamefont {Lee}},
  \bibinfo {author} {\bibfnamefont {I.~K.}\ \bibnamefont {Shmagin}}, \bibinfo
  {author} {\bibfnamefont {R.~M.}\ \bibnamefont {Kolbas}}, \bibinfo {author}
  {\bibfnamefont {H.~C.}\ \bibnamefont {Casey}}, \bibinfo {author}
  {\bibfnamefont {B.~P.}\ \bibnamefont {Keller}}, \bibinfo {author}
  {\bibfnamefont {U.~K.}\ \bibnamefont {Mishra}}, \ and\ \bibinfo {author}
  {\bibfnamefont {S.~P.}\ \bibnamefont {DenBaars}},\ }\href@noop {} {\bibfield
  {journal} {\bibinfo  {journal} {Appl. Phys. Lett.}\ }\textbf {\bibinfo
  {volume} {71}},\ \bibinfo {pages} {2572} (\bibinfo {year}
  {1997})}\BibitemShut {NoStop}%
\bibitem [{\citenamefont {Haensel}\ \emph {et~al.}(1969)\citenamefont
  {Haensel}, \citenamefont {Keitel}, \citenamefont {Koch}, \citenamefont
  {Skibowski},\ and\ \citenamefont {Schreiber}}]{HKKS69}%
  \BibitemOpen
  \bibfield  {author} {\bibinfo {author} {\bibfnamefont {R.}~\bibnamefont
  {Haensel}}, \bibinfo {author} {\bibfnamefont {G.}~\bibnamefont {Keitel}},
  \bibinfo {author} {\bibfnamefont {E.~E.}\ \bibnamefont {Koch}}, \bibinfo
  {author} {\bibfnamefont {M.}~\bibnamefont {Skibowski}}, \ and\ \bibinfo
  {author} {\bibfnamefont {P.}~\bibnamefont {Schreiber}},\ }\href@noop {}
  {\bibfield  {journal} {\bibinfo  {journal} {Phys. Rev. Lett.}\ }\textbf
  {\bibinfo {volume} {23}},\ \bibinfo {pages} {1160} (\bibinfo {year}
  {1969})}\BibitemShut {NoStop}%
\bibitem [{\citenamefont {Roessler}\ and\ \citenamefont
  {Walker}(1967{\natexlab{a}})}]{RW67}%
  \BibitemOpen
  \bibfield  {author} {\bibinfo {author} {\bibfnamefont {D.~M.}\ \bibnamefont
  {Roessler}}\ and\ \bibinfo {author} {\bibfnamefont {W.~C.}\ \bibnamefont
  {Walker}},\ }\href@noop {} {\bibfield  {journal} {\bibinfo  {journal} {J.
  Opt. Soc. Am.}\ }\textbf {\bibinfo {volume} {57}},\ \bibinfo {pages} {835}
  (\bibinfo {year} {1967}{\natexlab{a}})}\BibitemShut {NoStop}%
\bibitem [{\citenamefont {Saile}\ and\ \citenamefont {Koch}(1979)}]{SK79}%
  \BibitemOpen
  \bibfield  {author} {\bibinfo {author} {\bibfnamefont {V.}~\bibnamefont
  {Saile}}\ and\ \bibinfo {author} {\bibfnamefont {E.~E.}\ \bibnamefont
  {Koch}},\ }\href@noop {} {\bibfield  {journal} {\bibinfo  {journal} {Phys.
  Rev. B}\ }\textbf {\bibinfo {volume} {20}},\ \bibinfo {pages} {784} (\bibinfo
  {year} {1979})}\BibitemShut {NoStop}%
\bibitem [{\citenamefont {Leute}\ \emph {et~al.}(2009)\citenamefont {Leute},
  \citenamefont {Feneberg}, \citenamefont {Sauer}, \citenamefont {Thonke},
  \citenamefont {Thapa}, \citenamefont {Scholz}, \citenamefont {Taniyasu},\
  and\ \citenamefont {Kasu}}]{Leute09}%
  \BibitemOpen
  \bibfield  {author} {\bibinfo {author} {\bibfnamefont {R.~A.~R.}\
  \bibnamefont {Leute}}, \bibinfo {author} {\bibfnamefont {M.}~\bibnamefont
  {Feneberg}}, \bibinfo {author} {\bibfnamefont {R.}~\bibnamefont {Sauer}},
  \bibinfo {author} {\bibfnamefont {K.}~\bibnamefont {Thonke}}, \bibinfo
  {author} {\bibfnamefont {S.~B.}\ \bibnamefont {Thapa}}, \bibinfo {author}
  {\bibfnamefont {F.}~\bibnamefont {Scholz}}, \bibinfo {author} {\bibfnamefont
  {Y.}~\bibnamefont {Taniyasu}}, \ and\ \bibinfo {author} {\bibfnamefont
  {M.}~\bibnamefont {Kasu}},\ }\href@noop {} {\bibfield  {journal} {\bibinfo
  {journal} {Appl. Phys. Lett.}\ }\textbf {\bibinfo {volume} {95}},\ \bibinfo
  {pages} {031903} (\bibinfo {year} {2009})}\BibitemShut {NoStop}%
\bibitem [{\citenamefont {Roessler}\ and\ \citenamefont
  {Walker}(1967{\natexlab{b}})}]{Roessler67}%
  \BibitemOpen
  \bibfield  {author} {\bibinfo {author} {\bibfnamefont {D.~M.}\ \bibnamefont
  {Roessler}}\ and\ \bibinfo {author} {\bibfnamefont {W.~C.}\ \bibnamefont
  {Walker}},\ }\href@noop {} {\bibfield  {journal} {\bibinfo  {journal} {Phys.
  Rev.}\ }\textbf {\bibinfo {volume} {159}},\ \bibinfo {pages} {733} (\bibinfo
  {year} {1967}{\natexlab{b}})}\BibitemShut {NoStop}%
\bibitem [{\citenamefont {Ulbrich}(1985)}]{Ulbrich1985}%
  \BibitemOpen
  \bibfield  {author} {\bibinfo {author} {\bibfnamefont {R.~G.}\ \bibnamefont
  {Ulbrich}},\ }\href@noop {} {\bibfield  {journal} {\bibinfo  {journal} {Adv.
  Solid State Phys.}\ }\textbf {\bibinfo {volume} {25}},\ \bibinfo {pages}
  {299} (\bibinfo {year} {1985})}\BibitemShut {NoStop}%
\bibitem [{\citenamefont {Sottile}\ \emph {et~al.}(2007)\citenamefont
  {Sottile}, \citenamefont {Marsili}, \citenamefont {Olevano},\ and\
  \citenamefont {Reining}}]{Sottile07}%
  \BibitemOpen
  \bibfield  {author} {\bibinfo {author} {\bibfnamefont {F.}~\bibnamefont
  {Sottile}}, \bibinfo {author} {\bibfnamefont {M.}~\bibnamefont {Marsili}},
  \bibinfo {author} {\bibfnamefont {V.}~\bibnamefont {Olevano}}, \ and\
  \bibinfo {author} {\bibfnamefont {L.}~\bibnamefont {Reining}},\ }\href@noop
  {} {\bibfield  {journal} {\bibinfo  {journal} {Phys. Rev. B}\ }\textbf
  {\bibinfo {volume} {76}},\ \bibinfo {pages} {161103} (\bibinfo {year}
  {2007})}\BibitemShut {NoStop}%
\bibitem [{\citenamefont {Gori}\ \emph {et~al.}(2010)\citenamefont {Gori},
  \citenamefont {Rakel}, \citenamefont {Cobet}, \citenamefont {Richter},
  \citenamefont {Esser}, \citenamefont {Hoffmann}, \citenamefont {Del~Sole},
  \citenamefont {Cricenti},\ and\ \citenamefont {Pulci}}]{Gori10}%
  \BibitemOpen
  \bibfield  {author} {\bibinfo {author} {\bibfnamefont {P.}~\bibnamefont
  {Gori}}, \bibinfo {author} {\bibfnamefont {M.}~\bibnamefont {Rakel}},
  \bibinfo {author} {\bibfnamefont {C.}~\bibnamefont {Cobet}}, \bibinfo
  {author} {\bibfnamefont {W.}~\bibnamefont {Richter}}, \bibinfo {author}
  {\bibfnamefont {N.}~\bibnamefont {Esser}}, \bibinfo {author} {\bibfnamefont
  {A.}~\bibnamefont {Hoffmann}}, \bibinfo {author} {\bibfnamefont
  {R.}~\bibnamefont {Del~Sole}}, \bibinfo {author} {\bibfnamefont
  {A.}~\bibnamefont {Cricenti}}, \ and\ \bibinfo {author} {\bibfnamefont
  {O.}~\bibnamefont {Pulci}},\ }\href@noop {} {\bibfield  {journal} {\bibinfo
  {journal} {Phys. Rev. B}\ }\textbf {\bibinfo {volume} {81}},\ \bibinfo
  {pages} {125207} (\bibinfo {year} {2010})}\BibitemShut {NoStop}%
\end{thebibliography}%

\end{document}